\documentclass[10pt,twocolumn,letterpaper]{article}

\usepackage{iccv}
\usepackage{times}
\usepackage{epsfig}
\usepackage{graphicx}
\usepackage{amsmath}
\usepackage{bm}
\usepackage{bbm}
\usepackage{amssymb}
\usepackage{amsthm}
\usepackage{nicefrac}
\usepackage{algorithm}% http://ctan.org/pkg/algorithm
\usepackage{algpseudocode}% http://ctan.org/pkg/algorithmicx
\usepackage{siunitx}
\usepackage[dvipsnames]{xcolor}

\def\E{\mathbb E}
\def\P{\mathbb P}

\DeclareMathOperator*{\argmax}{arg\,max}
\DeclareMathOperator*{\argmin}{arg\,min}

\newcommand{\lam}{\Phi_\text{bkg}}
\newcommand{\muu}{\Phi_\text{sig}}
\newcommand{\td}{t_\text{d}}

\newtheorem{theorem}{Theorem}
\theoremstyle{theorem}\newtheorem{res}[theorem]{Result}
\theoremstyle{theorem}
\theoremstyle{definition}\newtheorem{defn}{Definition}
\theoremstyle{theorem}
\theoremstyle{remark}
\theoremstyle{remark}\newtheorem*{rmk*}{Remark}
\theoremstyle{definition}

\usepackage[pagebackref=true,breaklinks=true,colorlinks,bookmarks=false]{hyperref}
\newcommand*{\nolink}[1]{%
    {\protect\NoHyper#1\protect\endNoHyper}%
  }
\hypersetup{
    colorlinks = true,
    linkbordercolor={white},
    linkcolor=Cyan,
    urlcolor=Black,
    citecolor=Cyan
}

\iccvfinalcopy % *** Uncomment this line for the final submission

\def\iccvPaperID{1166} % *** Enter the ICCV Paper ID here

\newcommand{\mytitle}{\vspace{-0.2in}Asynchronous Single-Photon 3D Imaging}
\begin{document}

%%%%%%%%% TITLE
\title{\vspace{-15pt}\mytitle\vspace{-15pt}}

\author{
Anant Gupta* \,\,\,
Atul Ingle* \,\,\,
Mohit Gupta \\
{\tt \small \{anant,ingle,mohitg\}@cs.wisc.edu} \vspace{3pt}\\
University of Wisconsin-Madison \\
\framebox{\url{www.SinglePhoton3DImaging.com}}
}

\maketitle
\renewcommand*{\thefootnote}{$*$}
\setcounter{footnote}{1}
\footnotetext{Equal contribution}
\renewcommand*{\thefootnote}{$\dagger$}
\setcounter{footnote}{2}
\footnotetext{This research was supported by ONR grant
N00014-16-1-2995, DARPA REVEAL program and Wisconsin Alumni Research
Foundation.}
\renewcommand*{\thefootnote}{\arabic{footnote}}
\setcounter{footnote}{0}
\thispagestyle{empty}

%%%%%%%%% ABSTRACT
\begin{abstract}
%  Single-photon avalanche diodes (SPADs) are becoming popular for
%  time-of-flight depth-ranging (e.g., LiDAR) due to their unique ability to
%  capture individual photons with picosecond timing resolution. However,
%  ambient light (e.g., sunlight) incident on a SPAD LiDAR leads to severe
%  non-linear distortions (pileup) in the measured waveform, resulting in large
%  depth errors. We propose a family of acquisition schemes, called asynchronous
%  single-photon LiDAR, with the goal of mitigating pileup during data
%  acquisition itself.  Unlike the conventional synchronous operation,
%  asynchronous acquisition temporally misaligns SPAD measurement windows and
%  the laser cycles through deterministically predefined or randomized offsets.
%  Our key insight is that pileup distortions can be ``averaged out'' by
%  choosing a sequence of offsets that span the entire depth range. We derive a
%  generalized image formation model and perform theoretical analysis to explore
%  the space of asynchronous acquisition schemes, and design high-performance
%  schemes. Our simulations and experiments demonstrate an improvement in depth
%  accuracy of up to an order of magnitude as compared to the state-of-the-art,
%  across a wide range of high ambient light imaging scenarios.

Single-photon avalanche diodes (SPADs) are becoming popular in time-of-flight
depth-ranging due to their unique ability to capture individual
photons with picosecond timing resolution. However, ambient light (e.g.,
sunlight) incident on a SPAD-based 3D camera leads to severe non-linear distortions
(pileup) in the measured waveform, resulting in large depth errors. We propose
asynchronous single-photon 3D imaging, a family of acquisition schemes to
mitigate pileup during data acquisition itself. Asynchronous acquisition
temporally misaligns SPAD measurement windows and the laser cycles through
deterministically predefined or randomized offsets. Our key insight is that
pileup distortions can be ``averaged out'' by choosing a sequence of offsets
that span the entire depth range. We develop a generalized image formation
model and perform theoretical analysis to explore the space of asynchronous
acquisition schemes and design high-performance schemes. Our simulations and
experiments demonstrate an improvement in depth accuracy of up to an order of
magnitude as compared to the state-of-the-art, across a wide range of imaging
scenarios, including those with high ambient flux.
\end{abstract}

%LiDAR systems based on SPAD technology have the potential to provide much
%higher depth resolution at longer imaging range while operating at much lower
%laser source powers.
  
   %SPAD-based LiDAR systems have promised high resolution outdoor depth imaging
  %with high photon efficiency. However, the presence of strong ambient light
  %combined with large dead times leads to non-linear distortions, called
  %pileup, which adversely affects depth estimation. We propose a family of
  %optimal acquisition schemes that address the problem of pileup mitigate at
  %the data acquisition stage. Unlike the conventional synchronous mode of
  %operation, we show that pileup can be mitigated by operating the SPAD and
  %laser asynchronously to introduce arbitrary offsets to the measurement
  %windows. The combination of these two effects leads to a reduction in error
  %by an order of magnitude compared to the state-of-the-art methods. We
  %demonstrate the improvements using simulations and experiments, and provide a
  %glimpse of the rich space of computational techniques for SPAD based LiDAR.

%%%%%%%%% BODY TEXT
\section{Single-Photon Cameras} %<<<<< add more intuition, ``looser'' wordings here - seems too cramped right now.

% Pose the contribution as a thorough analysis of optimal acquisition schemes
% for SPAD-based LiDARs. Then make in Implications paragraph and talk about how
% our optimal acquisition scheme mitigates pileup and will enables LiDARs to
% work in high ambient light, and also chooses dead time optimally.

\begin{figure}
\centering \includegraphics[width=0.95\columnwidth]{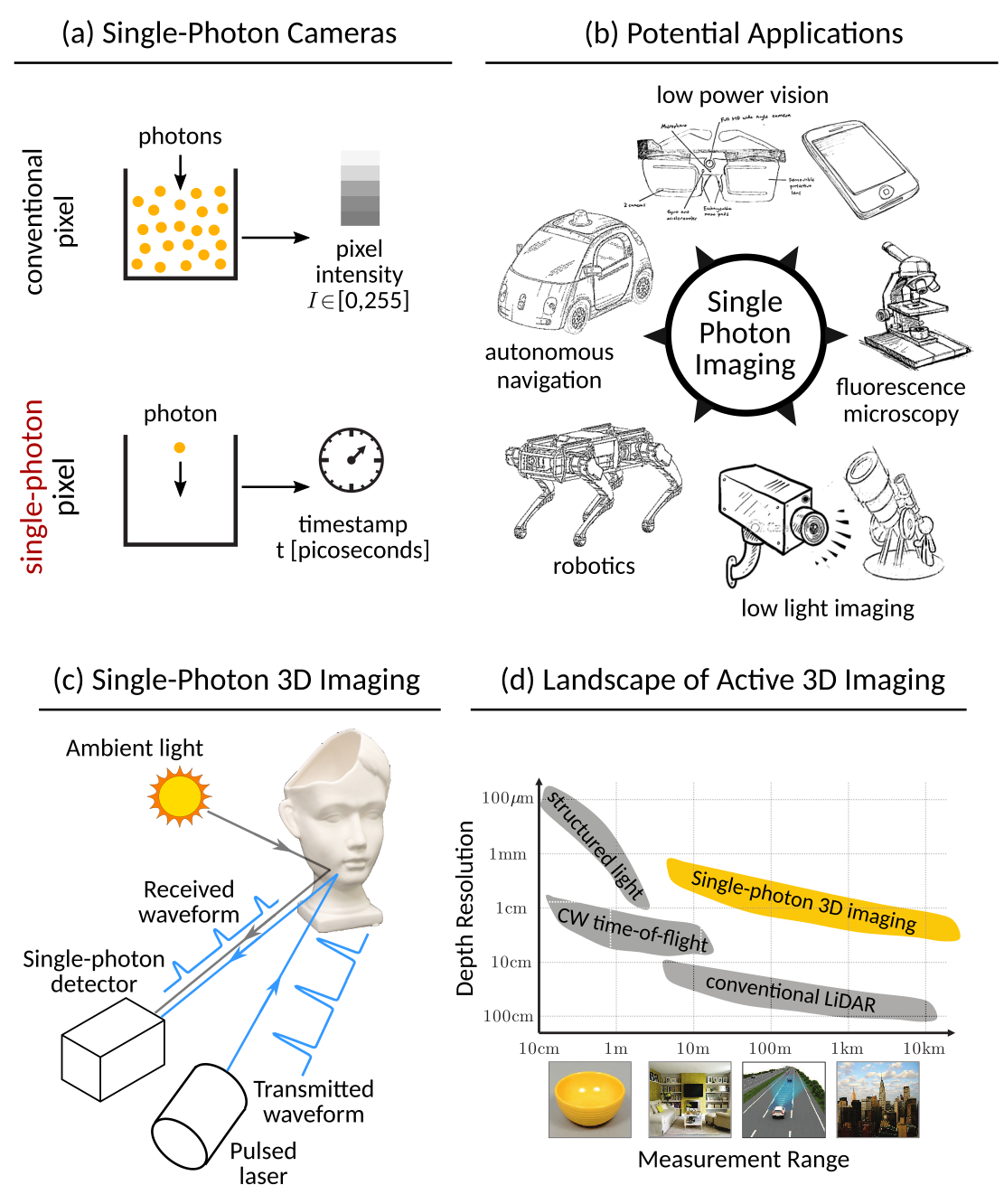}
\caption{ \small{{\bf Single-photon cameras and 3D imaging.}
(a) A single-photon camera pixel is sensitive to individual photons and can capture photon arrival times with picosecond resolution. (b) The extreme
sensitivity and resolution makes single-photon cameras promising candidates for several applications. (c) A single-photon 3D camera based on time-of-flight consists of a pulsed laser and a single-photon detector that timestamps returning photons. (d) Single-photon 3D cameras have the potential to provide extremely high depth resolution, even at long ranges.}}
\label{fig:teaser0}
\vspace{-0.2in}
\end{figure}

%are a fast emerging sensor technology. Unlike conventional photodiodes that produce an analog signal proportional to the incident photon flux, 

\begin{figure*}
  \includegraphics[width=1.0\textwidth]{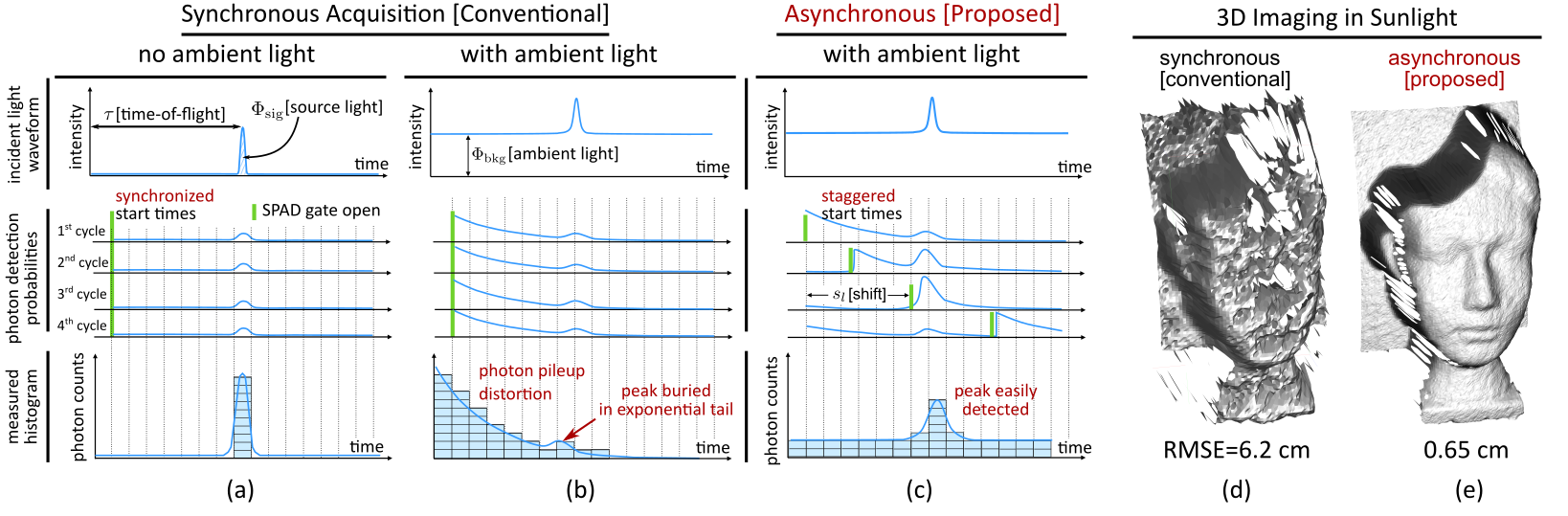}
  \caption{ {\bf Imaging model of single-photon 3D cameras.} (a) A single-photon 3D camera records the timestamps of returning photons over many laser cycles and constructs a histogram of photon arrival times. In the absence of ambient light, the peak of this histogram corresponds to the true depth. (b) In the conventional (synchronous) operation, ambient light causes photon pileup which distorts the histogram towards earlier time bins.
  (c) Asynchronous acquisition prevents pileup by temporally staggering the SPAD cycles with respect to the laser cycles, distributing the effect of pileup uniformly over all histogram bins. (d) 3D shape recovered using synchronous acquisition shows large depth errors due to pileup. (e) Proposed asynchronous method recovers accurate 3D shape even in high ambient light.
  \label{fig:teaser}} \vspace{-0.15in} 
\end{figure*}

Light is fundamentally quantized; any camera records incoming light not continuously, but in discrete packets called photons. A conventional camera typically captures hundreds to thousands of photons per pixel to create an image. What if cameras could record \emph{individual} photons, and, precisely measure their time-of-arrival? Not only would such cameras have extremely high sensitivity, but the captured data will have an additional time-dimension, a rich source of information inaccessible to conventional cameras. 

There is an emerging class of sensors, called single-photon avalanche diodes (SPADs)~\cite{rochas_spad_phd_2003} that promise single-photon
sensitivity (Fig.~\ref{fig:teaser0}(a)) and the ability to time-tag photons with picosecond precision. Due to these capabilities, SPADs are driving novel functionalities such as non-line-of-sight (NLOS) imaging~\cite{buttafava2015non,OToole2018} and microscopy of bio-phenomena at nano time-scales~\cite{Bruschini2019}. However, so far, SPADs are considered specialized devices suitable only for photon-starved (dark) scenarios, and thus, restricted to a limited set of niche applications. This raises the following questions: Can SPADs operate not just in low-light, but across the entire gamut of imaging conditions, including high-flux scenes \cite{Ingle_CVPR_2019}? In general, is it possible to leverage the exciting capabilities of SPADs for a broader set of mainstream computer vision applications (Fig.~\ref{fig:teaser0}(b))? 

In this paper, we address the above questions in the context of 3D imaging. Consider a single-photon 3D camera based on time-of-flight (ToF). It consists of a pulsed laser emitting periodic pulses of light toward the scene, and a SPAD sensor (Fig.~\ref{fig:teaser0}(c)). Although several conventional 3D cameras also use the ToF principle, single-photon 3D cameras have a fundamentally different imaging model. The SPAD detects \emph{at most one returning photon} per laser pulse, and records its time-of-arrival. Arrival times over several laser pulses are recorded to create a temporal histogram of photon arrivals, as shown in Fig.~\ref{fig:teaser}(a). Under low incident flux, the histogram is approximately a linearly scaled replica of the incident waveform, and thus, can be used to recover scene depths~\cite{Rapp,Lindell}. Due to the high timing resolution of SPADs, single-photon 3D cameras are capable of achieving ``laser-scan quality'' depth resolution (1--10 mm), at long distances (100--1000 meters) (Fig.~\ref{fig:teaser0}(d)). \smallskip

\noindent {\bf Single-photon 3D imaging in sunlight:} Due to the peculiar histogram formation process, single-photon 3D cameras cannot operate reliably under ambient light (e.g., sunlight in outdoor conditions). This is because early arriving ambient photons prevent the SPAD from measuring the signal (laser) photons that may arrive at a later time bin of the histogram. This distorts the histogram measurements towards earlier time bins, as shown in Fig.~\ref{fig:teaser}(b). This non-linear distortion, known as \emph{photon pileup}~\cite{Heide:2018:subpicosecond,pediredla2018signal,Kapusta:2015:TCSPC} makes it challenging to reliably locate the laser pulse, resulting in large depth errors. Although there has been a lot of research toward correcting these distortions in post-processing~\cite{Heide:2018:subpicosecond,Cominelli_2017,patting2018fluorescence,pediredla2018signal,rapp2018dead}, strong pileup due to ambient light continues to limit the scope of this otherwise exciting technology. 

%Although SPAD-based LiDARs hold considerable promise due to their single-photon sensitivity and extremely high timing (hence, depth) resolution, their image formation model as described above is fundamentally different as compared to conventional sensors (that directly capture the incident temporal waveform). 

%Consider a SPAD LiDAR operating outdoor in sunlight. 

%Single photon avalanche diodes (SPADs) can detect individual incident photons with high timing resolution, making them an attractive candidate for extreme imaging applications~\cite{Kirmani58,Pawlikowska_2017,Buttafava:15,OToole2018} including long-range LiDAR. 
%A SPAD-based LiDAR operates on the principle of time-correlated single photon counting (TCSPC). 

%TCSPC~\cite{becker2015advanced,Wahl_technote}, SPAD acquisition window is aligned with the emission of laser pulses. 

We propose asynchronous single-photon 3D imaging, a family of computational imaging techniques for SPAD-based 3D cameras with the goal of \emph{preventing} pileup during acquisition itself. In conventional ToF cameras, the laser and sensor are temporally synchronized. In contrast, we \emph{desynchronize} the SPAD acquisition windows with respect to the laser pulses. This introduces different temporal offsets between laser cycles and SPAD acquisition windows, as shown in Fig.~\ref{fig:teaser}(c). The key insight is that cycling through a range of temporal offsets (across different laser cycles) enables detecting photons in later time bins that would otherwise have been masked by early-arriving ambient photons. This distributes the effect of pileup across all histogram bins, thus eliminating the structured distortions caused by the synchronous measurements, as shown in Fig.~\ref{fig:teaser}(c).

At first glance, it may appear that such asynchronous measurements may not provide consistent depth information. The main idea lies in \emph{computationally re-synchronizing} the photon timing measurements with the laser cycles. To this end, we develop a generalized image formation model and derive a maximum likelihood estimator (MLE) of the true depth that accounts for arbitrary temporal offsets between measurement and laser cycles. Based on these ideas, we propose two asynchronous acquisition methods: uniform and photon-driven, which shift the SPAD window with respect to laser either deterministically or stochastically. These techniques can be implemented with minimal modifications to existing systems, while achieving up to an order-of-magnitude improvements in depth accuracy. An example is shown in Fig.~\ref{fig:teaser}(d--e).\smallskip

%; in the latter, randomness in photon arrival is leveraged for achieving a range of shifts. \smallskip

%We provide practical guidelines for optimizing  asynchronous acquisition, including combinations with complementary pileup mitigation approaches~\cite{Gupta_2019}.

%We perform  theoretical analysis, simulations and hardware experiments to demonstrate the performance benefits of the proposed approaches over conventional, synchronous acquisition. Furthermore, w

\noindent{\bf Implications and future outlook:} Due to their compatibility with
mainstream CMOS sensor fabrication lines, the capabilities of SPAD cameras
continue to grow
rapidly~\cite{Dutton_2016,Ulku_2019,gnanasambandam2019megapixel,ma2017photon,lee2018progress,Acconcia:2018:pileup,Beer_2018}.
As a result, the proposed methods, aided by rapid ongoing advances in SPAD
technology, will potentially spur wide-spread adoption of single-photon sensors
as \emph{all-purpose} cameras in demanding computer vision and robotics
applications, where the ability to perform reliably in both photon-starved and
photon-flooded scenarios is critical to success.

\section{Related Work}
\noindent{\bf Photon pileup mitigation for SPAD cameras:} Perhaps the most widely
adopted approach for preventing pileup is attenuation, i.e., optically blocking
the total photon flux incident on the SPAD so that only 1-5\% of the laser
pulses lead to a photon detection
\cite{becker2015advanced,Beer_2018}.\footnote{Note that attenuation blocks both ambient
and source photons. Attenuation can be achieved through various methods such as
spectral filtering, neutral density filtering or using an aperture stop.}
Recent work \cite{Heide:2018:subpicosecond,Gupta_2019} has shown that this
rule-of-thumb \emph{extreme attenuation} is too conservative and the optimal
operating flux is considerably higher. Various
computational~\cite{pediredla2018signal,Heide:2018:subpicosecond} and
hardware~\cite{Acconcia:2018:pileup,Beer_2018,Zhang_2018} techniques for
mitigating pileup have also been proposed. These approaches are complementary
to the proposed asynchronous acquisition, and can provide further improvements in
performance when used in combination.
\smallskip

\noindent{\bf Temporally shifted gated acquisition:} Fast-gated detectors
\cite{buttafava2014spad} have been used previously for range-gated LiDAR,
confocal microscopy and non-line-of-sight (NLOS) imaging \cite{buttafava2015non} to
preselect a specific depth range and suppress undesirable early-arriving
photons. A sequence of shifted SPAD gates has been used in
FLIM for improving temporal resolution and dynamic range
\cite{Wang_GatedCCD_1991,Tosi_FLIM_2011,Ulku_2019} and for extending
the unambiguous depth range of pulsed LiDARs \cite{Reilly_2014}. In contrast,
we use shifting to mitigate pileup and present a theoretically optimal method
for choosing the sequence of shifts and durations of the SPAD measurement gates
without any prior knowledge of scene depths.
\smallskip

\noindent{\bf Photon-driven acquisition:}
The photon-driven (or free-running) mode of operation has been analyzed for
FLIM~\cite{isbaner2016dead,Cominelli_2017,Acconcia:2018:pileup}, and recently
for LiDAR~\cite{rapp2018dead} where a Markov chain model-based iterative
optimization algorithm is proposed to recover the incident waveform from the
distorted histogram. The focus of these approaches is on designing efficient
waveform estimation algorithms. Our goal is different.  We explore the space of
asynchronous acquisition schemes with the aim of designing acquisition
strategies that mitigate depth errors due to pileup in high ambient light
under practical constraints such as a fixed time budget. We also propose a
generalized closed-form maximum likelihood estimator (MLE) for asynchronous
acquisition that can be computed without any iterative optimization routine.

% Our contribution is presenting a unified analysis of optimal acquisition
% schemes, we are not the first ones to propose free running acquisition.

%However, they operate in the low flux regime, where ambient flux is limited to
%at most 3 photons per cycle, and SBR is also high so that pile-up is due to
%both ambient and source. In contrast, we consider a much more challenging
%situation where ambient flux is upto 2 orders of magnitude higher (300 photons
%per cycle), and pile-up due to ambient flux dominates.
% include Isbaner et al paper

%\noindent{\bf Image formation model in shifting}: People have looked at the problem of
%modeling the SPAD measurements in the asynchronous mode, and recovering depth
%estimates \cite{rapp2018dead} or fluorescence lifetime estimates
%\cite{isbaner2016dead}. However, their models rely on incomplete sensor
%information to recover the unknown waveform. They model the distribution of the
%histogram of photon arrival times, and use complex filter-based approaches to
%disentangle the information about the waveform from the correlated histogram
%measurements. We provide the first probabilistic forward model that captures
%the likelihood of all the information available, and use it to compute the MLE.
%This can be interpreted as a generalization of the Coates's method
%\cite{Coates} to asynchronous systems. This interpretation lends itself to an
%elegant analysis of the depth estimation performance of asynchronous systems,
%and motivates the design of improved image acquisition schemes.

\section{Single-Photon 3D Imaging Model}\label{sec:Background}
A SPAD-based 3D camera consists of a pulsed laser that emits short periodic pulses
of light toward a scene point, and a co-located SPAD sensor that captures
the reflected photons (Fig.~\ref{fig:teaser0}(c)). Although the incident photon flux is a continuously varying function of time, a
SPAD has limited time resolution, resulting in a discrete sampling of the
continuous waveform. Let $\Delta$ denote the size of each discrete temporal bin
(usually on the order of few tens of picoseconds). Assuming an ideal laser
pulse modeled as a Dirac-delta function $\delta (t)$, the number of photons
incident on the SPAD in the $i^\text{th}$ time bin follows a Poisson
distribution with a mean given by:
\begin{equation}
  r_i = \muu \delta_{i,\tau} + \lam \,, \label{eq:true_waveform}
\end{equation}
where $\delta_{i,j}$ is the Kronecker delta,\footnote{$\delta_{i,j}=1$ for
$i=j$ and $0$ otherwise.} $\tau = \lfloor\nicefrac{2z}{c\Delta}\rfloor$ is the
discretized round-trip time delay, $z$ is the distance of the scene point from
the camera, and $c$ is the speed of light. $\muu$ is the mean number of signal
photons (due to the laser pulse) received per bin, and $\lam$ is the (undesirable)
background and dark count photon flux per bin. $B$ is the number of
time bins in a single laser period. The vector $\left(r_i\right)_{i=1}^B$ denotes the
\emph{incident photon flux waveform}. A reliable estimate of this waveform is needed
to estimate scene depth. \smallskip

\noindent{\bf Synchronous acquisition:} In order to estimate the incident waveform, SPAD-based 3D cameras employ the principle of time-correlated single-photon counting (TCSPC)~\cite{OToole:2017:SPAD,Kapusta:2015:TCSPC,becker2015advanced,Pellegrini:2000:Distance,Pawlikowska_2017,pediredla2018signal}. In conventional \emph{synchronous} acquisition, the SPAD starts acquiring photons immediately after the laser pulse is transmitted, as shown in Fig.~\ref{fig:teaser}(a). In each laser cycle (laser repetition period), after detecting the first incident photon, the SPAD enters a \emph{dead time} (${\sim}$\SI{100}{\nano\second}) during which it cannot detect additional photons. The SPAD may remain inactive for longer than the dead time so that the next SPAD acquisition window aligns with the next laser cycle.\footnote{The laser repetition period is set to $2z_\text{max}/c$, where $z_\text{max}$ is the unambiguous depth range. The photon flux is assumed to be 1-5\% \cite{Wahl_technote} of the laser repetition rate so that the probability of detecting photons in consecutive laser cycles is negligible. In high ambient light, the dead time from one cycle may extend into the next causing some cycles to be skipped.} The time of arrival of the first incident photon is recorded with respect to the start of the most recent cycle. A histogram $(N_1,\!\ldots,\!N_{B})$ of the first photon arrival times is constructed over many cycles, where $N_i$ denotes the number of times the first photon arrives in the $i^\text{th}$ bin. In low ambient light, the histogram, on average, is simply a scaled version of the incident waveform~\cite{Gupta_2019}, from which, depth can be estimated by locating its peak. \smallskip

\noindent {\bf Effect of ambient light in synchronous acquisition:} Under ambient light, the incident
flux waveform can be modeled as an impulse with a constant d.c. offset, as shown in the top of
Fig.~\ref{fig:teaser}(b). In high ambient flux, the SPAD detects an
ambient photon in the earlier histogram bins with high probability. This skews
the measured histogram towards earlier histogram bins, as shown in the bottom
of Fig.~\ref{fig:teaser}(b). The peak due to the laser source appears only as a
small blip in the exponentially decaying tail of the measured histogram. This
distortion, called photon pileup~\cite{Coates,becker2015advanced,pediredla2018signal},
significantly lowers the accuracy of depth estimates.

In the next two sections we introduce a generalization of the synchronous TCSPC
acquisition scheme, and show how it can be used to mitigate pileup distortion
and reliably estimate depths, even in the presence of high ambient light.

%because the bin corresponding to the true depth no longer receives the maximum
%number of photons. 

%In the extreme case, the later histogram bins might receive no photons, making
%depth reconstruction at those bins impossible.

%The measured histogram, however, does not reliably reproduce this ``DC shift''
%due to the peculiar histogram formation procedure that only captures the first
%photon for each laser cycle. 

%The number of histogram bins $B$ depends on the laser repetition period which
%is chosen based on the desired unambiguous depth range
%$z_\text{max}$.\footnote{} $B$ must satisfy $B\Delta \geq
%\nicefrac{2z_{\text{max}}}{c}$ to prevent distance aliasing.

\section{Theory of Asynchronous Image Formation \label{sec:asynchronous}}
In this section we develop a theoretical model for asynchronous single-photon 
3D cameras. We derive a histogram formation model and a generalized Coates's
estimator \cite{Coates} for the incident photon flux waveform.  In asynchronous
acquisition, we decouple the SPAD on/off times from the laser cycles
by allowing the SPAD acquisition windows to have arbitrary start times with
respect to the laser pulses (Fig.~\ref{fig:teaser}(c)).  A \emph{SPAD cycle} is
defined as the duration between two consecutive time instants when the SPAD
sensor is turned on. The SPAD will remain inactive during some portion of each
SPAD cycle due to its dead time.

%This is illustrated in Fig.~\ref{fig:teaser}(c).  We derive a histogram
%formation model and a generalized Coates-like \cite{Coates} estimator for the
%incident photon flux waveform for asynchronous acquisition. 
%
%In stark departure from the conventional TCSPC acquisition mode where the SPAD
%starts acquiring photons immediately after the laser pulse is transmitted, we
%operate the SPAD \emph{asynchronously} and allow the SPAD measurement windows
%to shift arbitrarily with respect to the laser pulse period. 

%The SPAD acquisition can be started at an arbitrary start time with respect to
%the most recent laser cycle. Moreover, the duration for which the SPAD remains
%on can also be varied programmatically. %Unlike the conventional TCSPC mode of
%operation, we assume that the SPAD acquisition window does not begin
%immediately with the transmission of the laser pulse. As shown in
%Fig.~\ref{fig:teaser}(c), the absolute timing of the histogram bins is
%dictated by the synchronization clock signal associated with the laser
%repetition frequency.  However, the start of each SPAD measurement window is
%arbitrarily shifted with respect to each laser pulse and hence maps to a
%different depth bin depending on the temporal shift. A subsequent laser pulse
%causes the SPAD measurement window to wrap around at the $B^\text{th}$
%histogram bin.
\begin{figure}[!ht]
\centering \includegraphics[width=1.0\columnwidth]{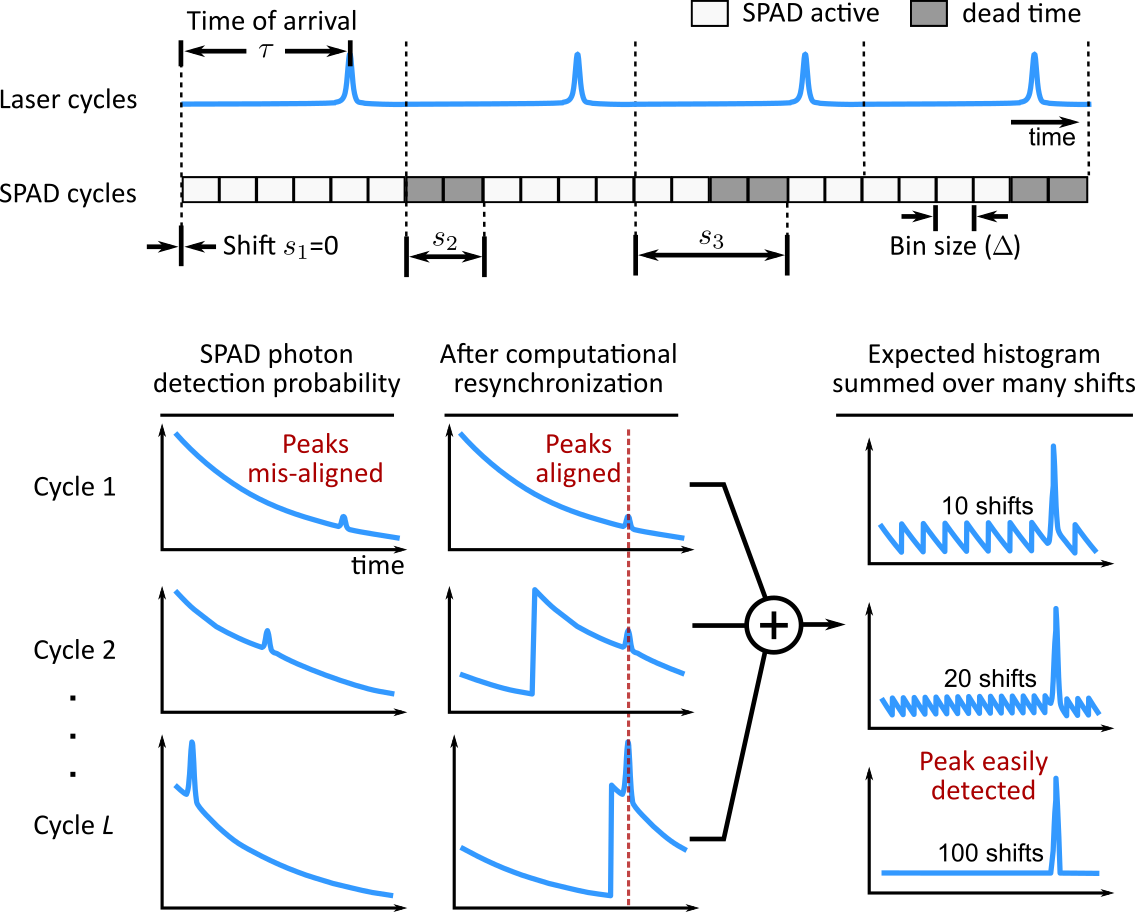}
\caption{ {\bf Histogram formation for asynchronous aquisition.} {\bf (Top)} The temporal location of the laser peak in the incident waveform corresponds to the round-trip time-of-flight. A slightly longer SPAD cycle period results in a sequence of increasing shifts with respect to the laser cycles. {\bf (Bottom)} The histogram formation process involves computational resynchronization of photon arrival times to the laser cycle boundaries, causing a ``wrap around.'' The measured histogram approaches the true waveform shape when a large number of uniformly spaced shifts is used. 
\label{fig:deterministic_shifting_intuition}}
\end{figure}

Each laser cycle consists of $B$ time bins which are used to build a photon count
histogram. The bin indices are defined with respect to the start of
the laser cycle, i.e., the first time bin is aligned with the transmission of
each laser pulse.  We assume that the laser repetition period is $B\Delta = 2
z_{\text{max}}/c$.  This ensures that a photon detected by the SPAD always
corresponds to an unambiguous depth range $[0, z_{\text{max}})$. Let  $s_l$ ($0
\leq s_l \leq B-1$) denote the bin index (with respect to the most recent laser cycle) at which the SPAD gate is activated
during the $l^\text{th}$ SPAD cycle ($1 \leq l \leq L$). As shown in
Fig.~\ref{fig:deterministic_shifting_intuition}(top), a SPAD cycle may extend
over multiple consecutive laser cycles. \smallskip

\noindent {\bf Probability distribution of measured histogram:}
Due to Poisson statistics, the probability $q_i$ that at least one photon is
\emph{incident} on the SPAD in the $i^\text{th}$ bin is:
\begin{equation}
  q_i = 1-e^{-r_i}, \label{eq:qi_and_ri}
\end{equation}
where $r_i$ is given by Eq.~(\ref{eq:true_waveform}).  A photon
\emph{detection} in the $i^\text{th}$ time bin occurs when no photon is
incident in the time bins preceding the $i^\text{th}$ bin in the current cycle,
and at least one photon is incident in the $i^\text{th}$ bin. The probability
$p_{l,i}$ of a photon detection in the $i^\text{th}$ bin in the $l^\text{th}$
SPAD cycle depends on the shift $s_l$, and is given by:
\begin{align}
p_{l,i} \hspace{0.05in} &= \hspace{0.05in} q_i \hspace{0.05in} \prod_{j: j<i} (1-q_j) \,,
%   p_{l,i} \hspace{0.05in} &= \hspace{0.05in} q_i \prod_{j \in J_{l,i}} (1-q_j)
%   \hspace{0.05in} = \hspace{0.05in} \left(1 - e^{-r_i}\right)\, e^{
%     -\!\!\sum\limits_{j \in J_{l,i}}\!\! r_j } \,,
    \label{eq:photon_detection_prob}
\end{align}
where it is understood (see \nolink{\ref{sn:async_image_formation}}) that
$j<i$ denotes the bin indexes preceding the $i^\text{th}$ bin in a modulo-B
sense with a ``wrap around'' depending on the shift $s_l$
(Fig.~\ref{fig:deterministic_shifting_intuition}(bottom)).  We introduce an
additional $(B\!+\!1)^\text{th}$ bin in the histogram to record the number of
cycles where no photons were detected, with corresponding bin probability
$p_{l,B+1}\! :=\! 1\! -\!\sum_{i=1}^B p_{l,i}$.

%For the $l^\text{th}$ laser cycle we define a one-hot random vector $(O_{l,1},
%O_{l,2},\ldots,O_{l,B},O_{l,B+1})$ as follows that marks the bin index where
%the photon was recorded.  Since the SPAD detects at most one photon per laser
%cycle, $(O_{l,i})_{i=1}^{B+1}$ contains zeros everywhere except at the bin
%index corresponding to the photon detection. Its joint distribution is given
%by:
%\[
%  (O_{l,i})_{i=1}^{B+1} \sim \text{($B\!+\!1$)-Categorical}\big((p_{l,i})_{i=1}^{B+1}\big).
%\]
%The final histogram of photon counts is obtained by summing these one-hot
%vectors over all laser cycles:
%\[ 
%  N_i = \sum_{l=1}^L O_{l,i} \, .
%\]

As in the synchronous case, we construct a histogram of the number of photons
detected in each time bin. Let $N_i$ be the number of photons captured in the
$i^\text{th}$ bin over $L$ SPAD cycles. As shown in
\nolink{\ref{sn:async_image_formation}}, the joint distribution of the measured
histogram $(N_1, N_2, \ldots, N_B, N_{B+1})$ is given by a Poisson-Multinomial
Distribution (PMD)~\cite{Daskalakis2015}.
The PMD is a generalization of the multinomial distribution; if $s_l = 0 \,
\forall \, l$ (conventional synchronous operation), this reduces to a multinomial
distribution~\cite{Heide:2018:subpicosecond,pediredla2018signal}.
\smallskip
%the MLE of photon incidence probabilities is given by:
%\begin{equation}
%  \widehat{q}_i = \frac{N_i}{D_i} \label{eq:estimated_incident_prob}
%\end{equation}

%The round-trip time-of-flight can be estimated simply by locating the peak in
%the histogram as $\widehat{\tau} = \argmax_{1\leq i \leq B} \widehat r_i$
%where $\widehat r_i \approx \nicefrac{N_i}{L}$. 

\noindent {\bf Characterizing pileup in asynchronous operation:}
Similar to the synchronous case, in the low incident flux regime ($r_i \ll 1 \,
\forall \, i$) the measured histogram is, on average, a linearly scaled version
of the incident flux: $\E[N_i] \approx L r_i$, and the incident flux can be
estimated as $\widehat{r}_i = \nicefrac{N_i}{L}$. However, in high ambient
light, the photon detection probability at a specific histogram bin depends on
its position with respect to the beginning of the SPAD cycle. Similar to
synchronous acquisition, histogram bins that are farther away from the start of
the SPAD cycle record photons with exponentially smaller probabilities compared
to those near the start of the cycle. However, unlike the synchronous case, the
shape of this pileup distortion wraps around at the $B^\text{th}$ histogram bin
during computational resynchronization. This is shown in
Fig.~\ref{fig:deterministic_shifting_intuition}(bottom). The segment that is
wrapped around depends on $s_l$ and may vary with each SPAD cycle.
\smallskip

%As a result, the measured
%histogram suffers from pileup distortions, albeit with two key differences with
%respect to synchronous LiDAR. 
%the photon counts over different histogram bins are correlated random
%variables with a joint distribution given by
%Eq.~(\ref{eq:histogram_joint_prob}). 

% First, the exponentially decaying shape of photon detection probabilities
% \emph{wraps around} at the $B^{th}$ histogram bin because the histogram bin
% indices are defined with respect to the start of the laser cycles. This can be
% seen mathematically through the ``modulo B'' wrap around in the bin indexes in
% Eq.~(\ref{eq:index_set}) and in Fig.~\ref{fig:deterministic_shifting_intuition}(e).
% Second, perhaps more importantly, the
% beginning index $s_l$ of the SPAD measurement window, and therefore the photon
% detection probabilities at a given histogram bin, could vary across laser
% cycles. This is depicted as the shifted locations of the exponentially decaying
% probabilities in Fig.~\ref{fig:teaser}(c) in different cycles. In contrast, in
% synchronous acquisition, the detection probabilities remain constant across
% cycles (Figure~\ref{fig:teaser} (b)). Therefore, in asynchronous acquisition,
% the (expected value of the) measured histogram is the sum of the photon
% detection probabilities over all the laser cycles. \smallskip

\noindent {\bf Computational pileup correction in asynchronous acquisition:}
A computational pileup correction algorithm must use the histogram
$(N_i)_{i=1}^{B+1}$ to estimate the true waveform $r_i$ via an estimate of
$q_i$ and Eq.~(\ref{eq:qi_and_ri}). Recall that a photon detection in a
specific histogram bin prevents subsequent bins from recording a photon.
Therefore, in the high flux regime, $q_i$ cannot be simply estimated as the
ratio of $N_i$ to the number of SPAD cycles ($L$); the denominator in this
ratio must account for the number of SPAD cycles where the $i^\text{th}$
histogram bin had an opportunity to record a photon.

\begin{defn}[{\bf Denominator Sequence}]
   Let $D_{l,i}$ be
an indicator random variable which is $1$ if, in the $l^\text{th}$ SPAD cycle,
no photon was detected before the $i^\text{th}$ time bin. The denominator sequence $(D_i)_{i=1}^B$ is defined as $D_i =  \sum_{l=1}^L
  D_{l,i}$.
  \label{def:den_seq}
\end{defn}

Note that $D_{l,i} = 1$ indicates that in the $l^\text{th}$ SPAD cycle, the
SPAD had an opportunity to detect a photon in the $i^\text{th}$ bin. By summing
over all SPAD cycles, $D_i$ denotes the total number of photon detection
opportunities in the $i^\text{th}$ histogram bin. Using this corrected
denominator, an estimate for $q_i$ is obtained as follows:
\begin{align*}
    \widehat{q}_i = \frac{N_i}{D_i} \,.
\end{align*}
% In addition to correcting pileup, the denominator sequence also serves as a
% measure of quality of the corrected estimate itself. Roughly speaking, $D_i$
% denotes the inverse of the variance of $\widehat{q_i}$.  The larger the value
% of $D_i$, the lower the variance of $\widehat{q}_i$, hence the more reliable
% the estimate.  Note that $D_i$ is itself a random variable, so we will be
% interested in its average or expected value. For a given incident flux
% waveform, the expected denominator sequence can be changed by changing the
% shifts used during acquisition. From now on, we refer to the expected
% denominator sequence as simply the denominator sequence.
We show in \nolink{\ref{sn:async_image_formation}} that $\widehat{q}_i$ is
in fact the MLE of $q_i$. The MLE of the incident flux
waveform is given by:
\begin{equation}
  \widehat{r}_i = \ln \left( \frac{1}{1 - \widehat{q}_i} \right)
  %= \ln \left( \frac{1}{1-\frac{N_i}{D_i}} \right)
  \label{eq:modified_coates_estimator}
\end{equation}
which is a generalization of the Coates's estimator
\cite{Coates,pediredla2018signal}.
Photon pileup causes later histogram bins to have $D_i \approx 0$ making it
difficult to estimate $r_i$.  Intuitively, a larger $D_i$ denotes more
``information'' in the $i^\text{th}$ bin, hence a more reliable estimate of the
true flux waveform can be obtained.
\smallskip

\vspace{-0.1in} 
% \section{Photon$\,$Pileup:$\,$Prevention$\,$Better$\,$than$\,$Cure?}
\section{Photon Pileup: Prevention Better than Cure?}
In theory, when operating in high ambient light, the generalized Coates's
estimator in Eq.~(\ref{eq:modified_coates_estimator}) can invert pileup
distortion for asynchronous acquisition with any given set of shifts $s_l$.
However, if the asynchronous acquisition scheme is not well-designed, this
inversion will lead to unreliable waveform estimates. For example, if the
shifts $s_l$ are all zero (synchronous acquisition), bins farther from the
start of the SPAD cycle will have $D_i \approx 0$ and suffer from extremely
noisy flux estimates. 

In this section, we design imaging techniques that prevent photon pileup in the acquisition phase itself, even under high ambient light. Our main observation is that delaying the start of the SPAD cycle with respect to the start of a laser cycle increases $D_i$ at later time bins. \emph{The key idea}, as shown in Fig.~\ref{fig:deterministic_shifting_intuition}, is to cycle through various shifts $s_l$ for different SPAD cycles. This ensures that each time bin is close to the start in at least a few SPAD cycles. Intuitively, if all possible shifts from $0$ to $B\!-\!1$ are used, the effect of the
exponentially decaying pileup due to ambient photons gets distributed over all histogram bins equally. On the other hand, returning signal photons from the true laser peak add up ``coherently'' because their bin location remains fixed. As a result, the accumulated histogram has enough photons in all bins
(Fig.~\ref{fig:deterministic_shifting_intuition}(e)) to enable reliable Coates's estimates.
% This suggests the following acquisition strategy for mitigating pileup:
% Gradually shift each SPAD cycle with respect to the laser cycle. 

\begin{figure}[!t]
  \centering \includegraphics[width=0.98\columnwidth]{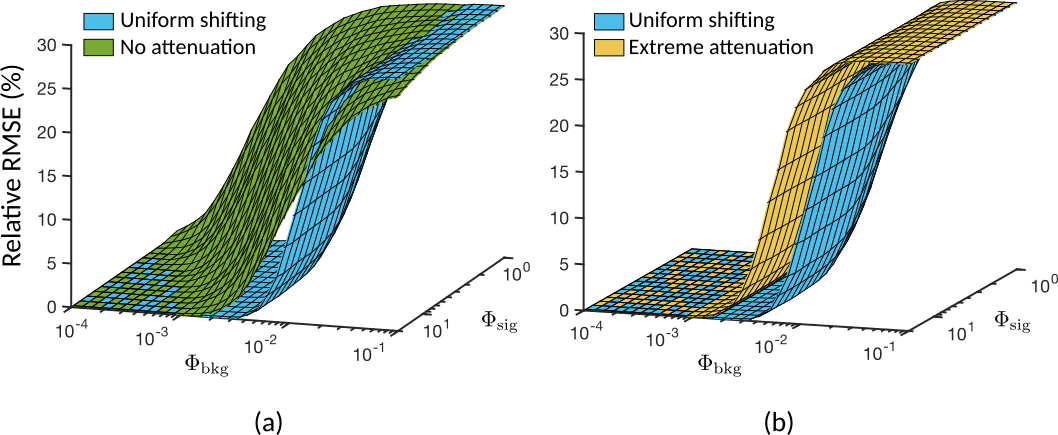}
  \caption{ {\bf Simulated depth RMSE at different ambient and signal flux
  levels.} Asynchronous acquisition with uniform shifting achieves lower error than synchronous acquisition with
  no and extreme attenuation~\cite{Gupta_2019}, over a wide range of flux conditions. \label{fig:tcspc_vs_determinsitic}}
  \vspace{-0.15in}
\end{figure}

We characterize the space of all shifting strategies by their \emph{shift
sequence}, $(s_i)_{i=1}^L$. For now, we only consider \emph{deterministic shift
sequences}, which means that the shifts are fixed and known prior to acquisition. Given these definitions, the question that we seek to address is: What is the optimal shifting strategy that minimizes depth estimation error?
% In addition to correcting pileup, the denominator sequence also serves as a
% measure of quality of the corrected estimate itself. Roughly speaking, $D_i$
% denotes the inverse of the variance of $\widehat{q_i}$.  The larger the value
% of $D_i$, the lower the variance of $\widehat{q}_i$, hence the more reliable
% the estimate.  Note that $D_i$ is itself a random variable, so we will be
% interested in its average or expected value. For a given incident flux
% waveform, the expected denominator sequence can be changed by changing the
% shifts used during acquisition. From now on, we refer to the expected
% denominator sequence as simply the denominator sequence.  We answer this
% question in two steps. First, we show that depth error can be expressed as a
% function of the expected denominator sequence. As a result, we can compare
% errors of different shifting strategies by simply comparing their denominator
% sequences.
% It turns out that we can express the optimality of a shifting strategy in terms
% of a certain property exhibited by its denominator sequence.
We now present two key theoretical results towards answering this question for
a SPAD-based 3D camera operating in the \emph{high ambient flux regime}
where the total number of incident photons is dominated by ambient photons.\footnote{We
  define signal-to-background-ratio $\mathsf{SBR}=\muu / B\lam$. In the high ambient flux regime, $\mathsf{SBR}\ll 1$.}

%The first result (based on \cite{Gupta_2019}) shows that if a shifting
%strategy can achieve a constant expected denominator sequence, it will have
%lower depth error (roughly speaking) than all other shifting strategies.  The
%second result shows that there exists a shifting strategy that achieves a
%constant expected denominator: \emph{uniform shifting}.

%\begin{defn}[{\bf High Ambient Flux Regime}]
%  A scenario where the total number of incident photons is dominated by ambient photons, i.e.:
%  $B\lam \gg \muu$, i.e., $\mathsf{SBR}\footnote{We
%  define $\mathsf{SBR}=\muu / B\lam$ as the ratio of the average number of
%laser photons to ambient photons incident in a laser cycle.} \ll 1$.
%\end{defn}

\begin{defn}[{\bf Uniform Shifting}]
A shifting strategy is said to be uniform if its shift sequence is a uniform partition of the time
interval $[0,B\Delta)$, i.e., is a permutation of the sequence
$\left(0, \lfloor\nicefrac{B}{L}\rfloor, \lfloor\nicefrac{2B}{L}\rfloor, ...,
\lfloor\nicefrac{(L-1)B}{L}\rfloor\right)$.
\end{defn}

\begin{res}[{\bf Denominator Sequence and Probability of Depth Error}]
  In the high ambient flux regime, among all denominator sequences with a fixed
  total expected sum $\sum_{i=1}^L \E[D_i]$, an upper bound on the average
  probability of depth error for the estimator in
  Eq.~(\ref{eq:modified_coates_estimator}) is minimized when $\E[D_i] = \E[D_j]
  \,\forall\, i,j$.
  \label{thm1}
\end{res}

\begin{res}[{\bf Denominator Sequence for Uniform Shifting}]
  Uniform shifting achieves a constant expected denominator
  sequence.
  \label{thm2}
\end{res}

\noindent{\bf Interpreting Results~\ref{thm1} and \ref{thm2}:} As shown in
\nolink{\ref{sn:res1_res2_proofs}}, for a fixed $L$, different shift sequences
will lead to different denominator sequences but the total expected denominator
$\sum_{i=1}^L \E[D_i]$ remains constant.  The first result (based on
\cite{Gupta_2019}) shows that if a shifting strategy can achieve a constant
expected denominator sequence, it will have lower depth error than all other
shifting strategies (including synchronous acquisition).  The second result
shows that there exists a shifting strategy that achieves a constant expected
denominator: uniform shifting.  As a byproduct, a uniform denominator sequence
makes the depth errors invariant to the true bin location, unlike the
synchronous case where later time bins suffer from higher depth errors.

%Therefore, Result~\ref{thm1} suggests
%that among all deterministic shifting strategies, including 
%the special case of $s_l=0 \forall l$ (synchronous acquisition), those that 
%provide a constant expected denominator sequence are desirable.
%Result~\ref{thm2} suggests that uniform shifting ensures that each histogram
%bin is close to the start of the SPAD cycle in an equal number of cycles. This
%provides an equal opportunity to each histogram bin to detect photons, thus
%resulting in a constant expected denominator sequence, and by Result
%\ref{thm1}, minimizes overall depth error. As a byproduct, a uniform
%denominator sequence makes the depth errors invariant to the true bin location,
%unlike the synchronous case where later time bins suffer from higher depth
%errors.
\smallskip

% \noindent {\bf $\bm{\ell_0}$ vs. $\bm{\ell_2}$ depth error:}
% Ultimately, our goal is to design shifting strategies that provide the best
% performance in a root-mean-squared error (RMSE) sense ($\ell_2$ depth error).
% Result~\ref{thm1} uses an upper bound on the probability of depth estimation
% error ($\ell_0$ error) as a surrogate for RMSE. This makes mathematical
% analysis tractable, enabling derivation of closed-form expressions. Our
% simulations and experimental results show that the resulting shifting
% strategies perform well in $\ell_2$ sense as well and achieve considerably
% lower RMSE than the state-of-the-art.
% \smallskip

\noindent{\bf Single-pixel simulations:} 
We compare the performance of uniform shifting and conventional synchronous
acquisition through Monte Carlo simulations.\footnote{
  The maximum possible relative RMSE is 30\% because it is defined in a modulo-B sense. See
\nolink{\ref{sn:sim_expt_details}}.}
We use a histogram with $B=1000$ and $\Delta=\SI{100}{\pico\second}$ and
consider a wide range of background and signal photon flux levels in the
discrete delta pulse model of Eq.~(\ref{eq:true_waveform}). Uniform
shifts are simulated by choosing equally spaced shifts between $0$ and
$1000$ and generating photon counts using Eq.~(\nolink{\ref{eq:histogram_joint_prob}}).
Depth is estimated using the generalized estimator
(Eq.~(\ref{eq:modified_coates_estimator})). As seen in
Fig.~\ref{fig:tcspc_vs_determinsitic}, the depth RMSE with uniform shifting is
considerably lower than conventional synchronous acquisition. At certain
combinations of signal and background flux levels, uniform shifting estimates
depths with almost zero RMSE while the conventional methods give a very high
error.

\section{Practically Optimal Acquisition for Single-Photon 3D Imaging in Bright
Sunlight} 

\begin{figure}[!t]
  \centering\includegraphics[width=1.0\columnwidth]{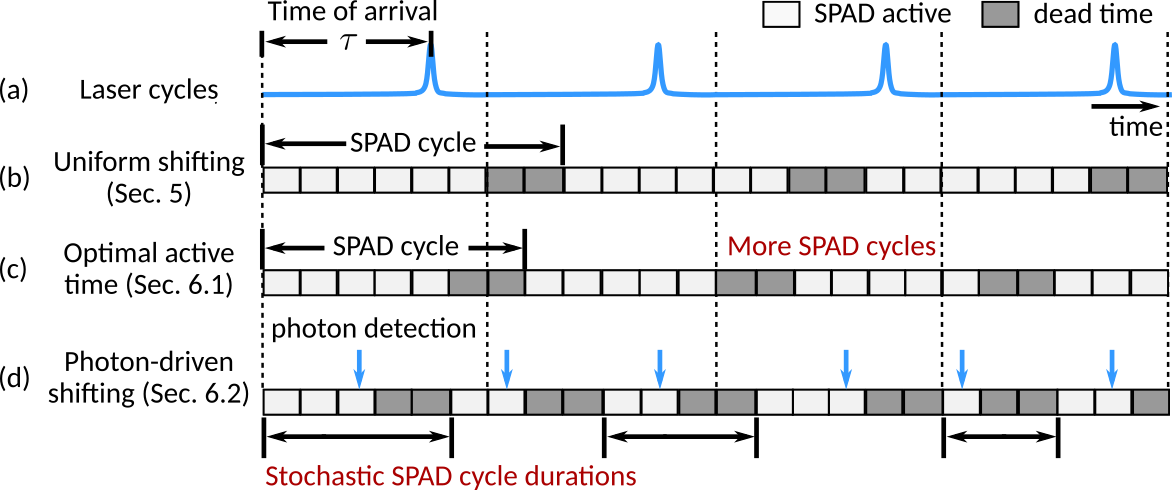}
  \caption{ {\bf Different asynchronous acquisition methods.} (a) The
  incident waveform has a period equal to the laser cycle period. (b)
  Uniform shifting staggers the laser and SPAD cycles by introducing a
  mismatch in cycle lengths. (c) Optimizing the SPAD active
  time enables more SPAD cycles to fit within a fixed total capture
  time. (d) Photon-driven shifting has random SPAD cycle lengths
  determined by photon detection events. 
  \label{fig:shifting_strategies}
  \vspace{-0.15in}
  }
\end{figure}

The theoretical analysis in the previous section shows uniform shifting
minimizes an upper bound on the $\ell_0$ depth error. It is natural to ask: How
can we implement practical uniform shifting approaches that are not just
theoretically optimal in the $\ell_0$ sense, but also achieve good RMSE
($\ell_2$ error) performance under realistic constraints and limited
acquisition time? In this section, we design several high-performance shifting
schemes based on uniform shifting. These are summarized in
Fig.~\ref{fig:shifting_strategies}.

Uniform shifting can be implemented in practice by making the SPAD cycle period
longer than the laser cycle (Fig.~\ref{fig:shifting_strategies}(b)), and
relying on this mismatch to automatically cycle through all possible shifts.
Moreover, this can be implemented at a negligible additional cost in terms of
total acquisition time as shown in \nolink{\ref{sn:fractional_shifts}}.

\subsection{SPAD Active Time Optimization}\label{sec:ActiveTimeOptimization}
So far, we have assumed the SPAD active time duration is fixed and equal to
$B\Delta$. Programmable fast-gated SPAD detectors
\cite{buttafava2014spad,Burri_2014} allow  flexibility in choosing different
active time and SPAD cycle durations (Fig.~\ref{fig:shifting_strategies}(c)).
Arbitrary shift sequences can also be implemented by varying the number of
active time bins, $m$, while keeping the inactive duration fixed at $\td$. This
expands the space of shifting strategies characterized by the active time bins,
$m$, and the shift sequence, $(s_l)_{l=1}^L$.  Under a fixed acquisition time
constraint:
\begin{equation} 
L (m\Delta + t_d) \leq T. \label{eq:acquisition_time}
\end{equation}
Note that $L$ can now vary with $m$. Can this greater design
flexibility be used to improve depth estimates?

\begin{figure}[!t]
  \centering\includegraphics[width=1.0\columnwidth]{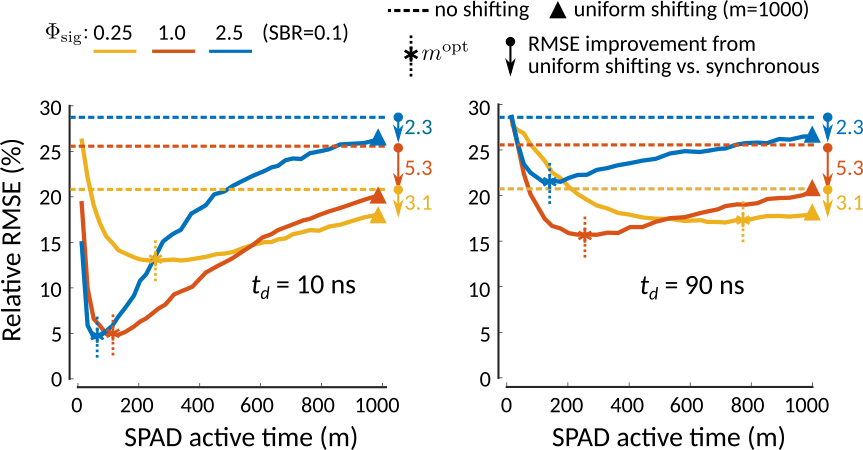}
  \caption{ {\bf Effect of SPAD active time on performance.} This plot shows the improvement in RMSE from SPAD active time optimization at
  different signal strengths and dead times. Note that the RMSE gain (size of vertical arrows) due to uniform shifting over synchronous acquisition remains unchanged across dead times. For low flux levels and long dead times, a longer active time improves RMSE by enabling the SPAD to capture more
  photons.
  \label{fig:error_vs_m}} \vspace{-0.15in}
\end{figure}

Varying $m$ leads to an interesting trade-off. Shortening the active time
duration causes a larger proportion of each SPAD cycle to be taken up by dead
time. On the other hand, using a very long active time is inefficient because
the portion of the active time after the first photon arrival is spent in dead
time anyway. This raises the question: What is the optimal active time that
minimizes the depth error? In \nolink{\ref{sn:mopt_derivation}} we show that
the optimal active time for uniform shifting is given by:
\begin{equation}
  m^\text{opt} = \argmax_m  \frac{T}{m\Delta+\td}\frac{1-e^{-m \lam}}{1-e^{-\lam}}.
  \label{eq:optimal_m}
\end{equation}

\smallskip

\noindent{\bf Simulation results for varying active time:}
Fig.~\ref{fig:error_vs_m} shows plots of depth RMSE vs. $m$ for a wide range of
ambient flux levels and two different values of dead time.  Observe that the
RMSE curves have  local minima which agree with our theory
(Eq.~(\ref{eq:optimal_m})). For a wide range of photon flux levels considered
here, $m^\text{opt}$ is \emph{shorter} than the conventionally used active time
of $m=B=1000$ and gives a remarkable improvement in RMSE by up to a factor of
\num{6}.

\subsection{Photon-Driven Shifting}\label{sec:PhotonDriven}
The optimal active time criterion balances the tradeoff between short and long
active time windows in an average sense. However, due to the large variance in
the arrival time of the first photon in each cycle, a fixed $m$ cannot achieve
both these goals on a per-photon basis. It is possible to achieve
photon-adaptive active time durations using the free-running
mode~\cite{rapp2018dead} where the SPAD is always active, except after a photon
detection when it enters a dead time. 

\begin{figure}[!t]
  \centering\includegraphics[width=1.0\columnwidth]{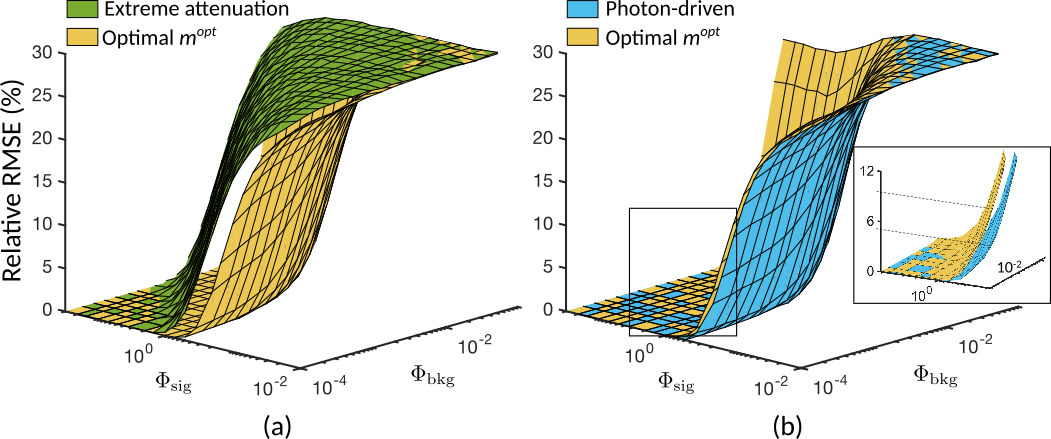}
  \caption{ {\bf Simulation-based evaluation of practically optimal asynchronous acquisition.}
  (a) Asynchronous acquisition with optimal SPAD active time (Section~\ref{sec:ActiveTimeOptimization}) provides an order
  of magnitude lower depth RMSE as compared to existing methods. (b) Photon-driven shifting (Section~\ref{sec:PhotonDriven}) further lowers RMSE by allowing the active time to vary stochastically on a per-photon basis. \label{fig:photon_driven_surfaceplot}}
  \vspace{-0.2in}
\end{figure}

In the free-running mode the active times and SPAD cycle durations
vary randomly due to the stochastic nature of photon arrivals. As shown in
Fig.~\ref{fig:shifting_strategies}(d), this creates different shifts
across different SPAD cycles. Over a sufficiently long acquisition time, a uniformly
spaced sequence of shifts is achieved with high probability, distributing the
effect of pileup over all histogram bins uniformly. We call this randomized
shifting phenomenon \emph{photon-driven shifting}.
\smallskip
% As for earlier shifting schemes, the denominator sequence plays an important
% part in the design and error analysis of depth estimators for photon-driven
% shifting.

\noindent{\bf Depth estimator for photon-driven shifting:}
Unlike deterministic shifting, the shift sequence in photon-driven shifting is
stochastic because it depends on the random photon arrival times. In
\nolink{\ref{sn:photon_driven_denominator}} we show that the scene depths can
still be estimated using the generalized Coates's estimator
(Eq.~(\ref{eq:modified_coates_estimator})) as before\footnote{Note that the
joint distribution of $(N_i)_{i=1}^L$ derived in
Section~\ref{sec:asynchronous} does not hold for photon-driven shifting
because the shift sequence is random.}.

% \noindent{\bf Theoretical Analysis Using Denominator Sequence:}
% As for earlier shifting schemes, we analyze the performance of photon-driven
% shifting via the expected denominator sequence.
The following result states
that photon-driving shifting possesses the desirable property of providing a
uniform shift sequence.  See
\nolink{\ref{sn:photon_driven_denominator}} for a proof.

\begin{res}
  As $L\rightarrow \infty$, photon-driven shifting achieves a uniform shift
  sequence.
%\[
%  \E[D_i] = \frac{T}{B(1+(1-e^{-\lam})\td)}, \,\,\,\,\,\,\,\, 1\leq i\leq B.
%\]
%for $1\leq i\leq B$. 
\label{thm3}
\end{res}

Result \ref{thm3} says that photon-driven shifting exhibits a pileup averaging
effect, similar to uniform shifting. Although this does not establish a
relationship between the shift sequence and depth RMSE, our results show up to
an order of magnitude improvement in RMSE compared to conventional synchronous
acquisition.\footnote{Proving exact uniformity of the shift sequence requires
assuming $L \to \infty$, although in practice, we found that $L \leq 50$ SPAD
cycles is sufficient.}
\smallskip

\begin{figure}[!t]
  \centering\includegraphics[width=0.98\columnwidth]{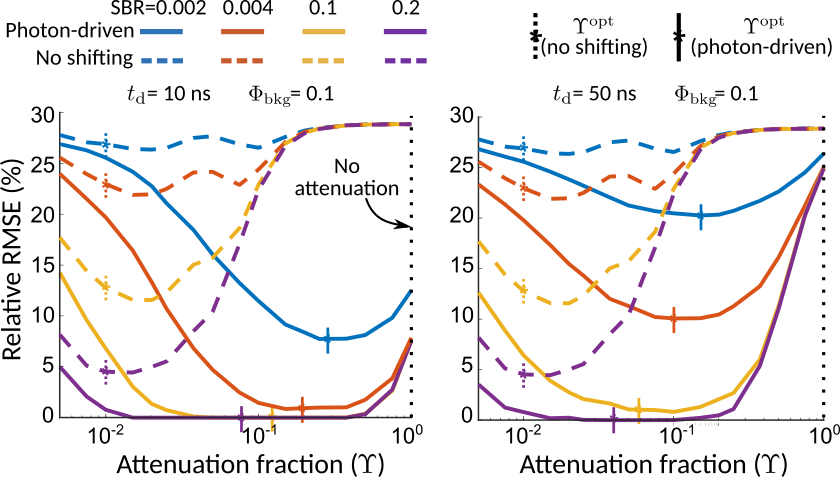}
  \caption{ {\bf Optimal attenuation factor $\Upsilon^\text{opt}$} for
  photon-driven shifting is higher than that of synchronous acquisition,
  leading to efficient flux utilization, while minimizing photon pileup. 
  \label{fig:free_running_vs_no_shifting_attenuation}}
\vspace{-0.15in}
\end{figure}

\begin{figure*}[!ht]
  \centering\includegraphics[width=0.9\textwidth]{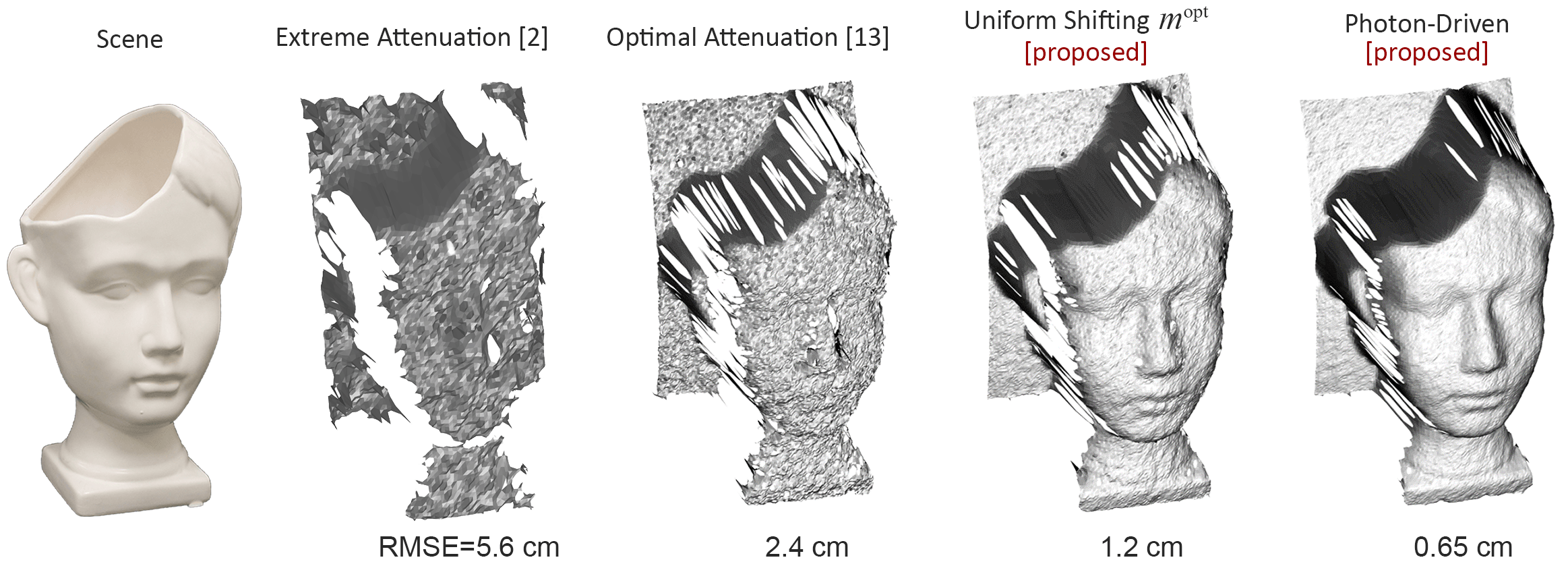}
  \caption{{\bf Experimental demonstration of single-photon 3D imaging under strong ambient light.} A white ``PorcelainFace'' vase was illuminated with high ambient light of $B\lam=11$ photons and scanned with a low-power laser at an SBR of \num{0.02}. The proposed asynchronous acquisition schemes achieve considerably higher depth quality as compared to conventional synchronous methods.
  \label{fig:face_result}}
  \vspace{-0.0in}
\end{figure*}

\begin{figure*}
  \centering\includegraphics[width=0.9\textwidth]{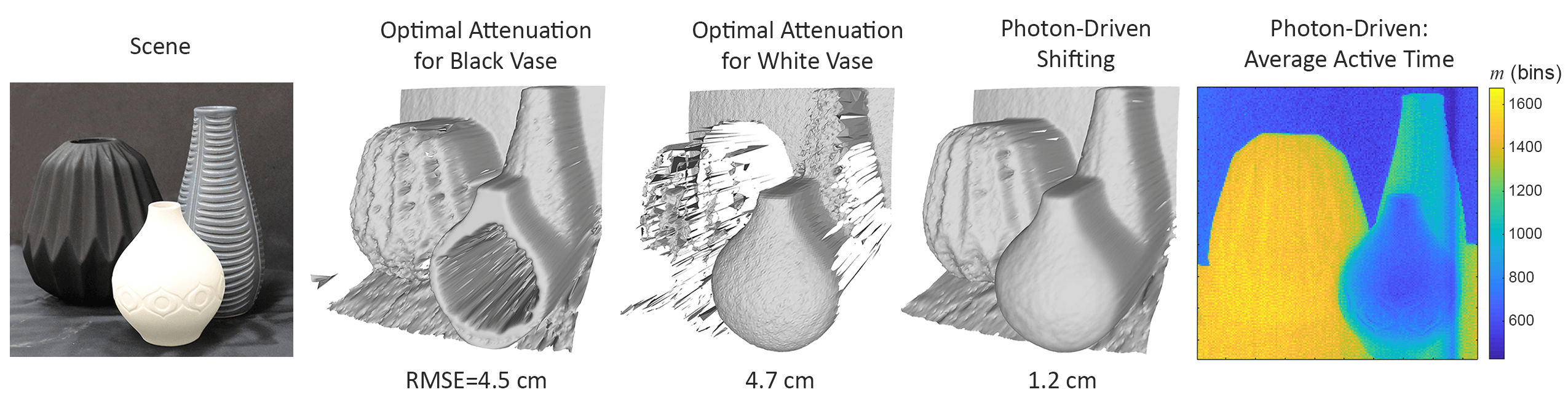}
  \caption{ {\bf Adaptivity of photon-driven shifting to different albedos.}
  The black vase in this ``Vases'' scene has $\nicefrac{1}{10}^\text{th}$ the reflectivity of the white vase. With synchronous acquisition, the attenuation fraction must be adjusted individually for each vase. In contrast, both vases are reliably reconstructed with photon-driven shifting which
  automatically adapts the active time duration for each pixel.\label{fig:vase_result}}
\vspace{-0.15in}
\end{figure*}

\noindent{\bf Simulation results:}
Fig.~\ref{fig:photon_driven_surfaceplot} shows simulated RMSE results for
photon-driven shifting over a wide range of signal and ambient flux levels.
For some flux levels the proposed shifting methods provide almost zero depth
error while the conventional method has the maximum possible error. The RMSE of
photon-driven shifting is similar to uniform shifting with $m^\text{opt}$,
but for some flux levels it can provide a factor of \num{2}
improvement over uniform shifting. \nolink{\ref{sn:free_vs_det}} discusses
certain regimes where deterministic shifting may be preferable over photon-driven
shifting.

\subsection{Combination with Flux Attenuation}
Recent work~\cite{Gupta_2019,Heide:2018:subpicosecond} has shown that there is
an optimal incident flux level at which pileup in a synchronous SPAD-based 3D
camera is minimized while maintaining high SNR. This optimal flux can be
achieved by optically attenuating the incident photon flux. In the space of
acquisition strategies to deal with pileup, attenuation can be considered a
complementary approach to asynchronous shifting. In
\nolink{\ref{sn:filtering_with_shifting}}, we show that the  optimal
attenuation fraction for photon-driven shifting is given by:
\[
  \Upsilon^\text{opt}_\text{photon-driven} = \min \left\{1.0, \argmin_\Upsilon \frac{1+\td(1-e^{-\Upsilon \lam})}{e^{-\Upsilon\lam}(1-e^{-\Upsilon\muu})} \right\}.
\]
Fig.~\ref{fig:free_running_vs_no_shifting_attenuation} shows simulation results
of depth RMSE for the conventional synchronous mode and photon-driven shifting
over a range of attenuation factors and two different dead times. The locations
of the minima agree with our theory. There are two key observations.  First,
the optimal attenuation fraction with shifting is much higher than that for
conventional synchronous acquisition. Second, combining attenuation with
photon-driven shifting can provide a large gain in depth error performance,
reducing the RMSE to almost zero under certain conditions.

\section{Experiments}
Our hardware prototype consists of a
\SI{405}{\nano\meter} pulsed laser (Picoquant LDH-P-C-405B), a TCSPC module
(Picoquant HydraHarp 400) and a fast-gated SPAD \cite{buttafava2014spad} that
can be operated in both triggered and free-running modes and has a
programmable dead time which was set to \SI{50}{\nano\second}. 
We operated the laser at a repetition frequency of
\SI{10}{\mega\hertz} for an unambiguous depth range of \SI{15}{\meter}
discretized into \num{1000} histogram bins. For uniform shifting, we operated the SPAD with its internal clock to obtain
shifts between the SPAD measurement windows and the laser cycles.

% \noindent{\bf Single-pixel results:} Fig.~\ref{fig:single_pixel_expt} shows
% depth RMSE at four different ambient flux levels for a range of SBR values. A
% combination of photon-driven shifting with optimal attenuation provides the
% best performance of all techniques. For high ambient flux levels, even at high
% SBR, the conventional 5\% rule-of-thumb and shifting alone fail to provide
% acceptable depth reconstruction performance. This suggests that the optimal
% acquisition strategy for SPAD-based LiDAR must use a combination of both
% shifting and attenuation.

\noindent{\bf 3D point-scanning results:} Fig.~\ref{fig:face_result} shows 3D
reconstructions of ``PorcelainFace'' scene under high ambient illumination.
Both uniform and photon-driven shifting
($\Upsilon^\text{opt}_\text{photon-driven}=1$) perform better than synchronous
acquisition methods. Photon-driven acquisition provides sub-centimeter RMSE,
which is an order of magnitude better than the state-of-the-art extreme
attenuation method.

The method of \cite{Gupta_2019} uses synchronous acquisition and relies on
setting an attenuation factor for different parts of the scene based on the
total photon flux and hence requires pixel-wise adaptation.  The ``Vases''
scene in Fig.~\ref{fig:vase_result} consists of a black vase with a much lower
albedo than the white vase.  The attenuation fraction needed for the white
vase is too low and causes the black vase to appear noisy, whereas the
attenuation fraction for the black vase is too high to avoid pileup distortions
at the white vase. The average active time with photon-driven shifting
($\Upsilon^\text{opt}_\text{photon-driven}=1$) automatically adapts to different
photon flux levels and reliably captures the depth map for both vases.  For
darker scene points the average active time is \emph{longer} than the laser
cycle period of $B=1000$.

%Note that for both scenes, no
%attenuation was needed for photon-driven shifting as the flux was already at or
%below optimal.

\section{Limitations and Discussion}
\noindent{\bf Incorporating spatial priors:}
The theoretical analysis and results presented here are limited to a pixel-wise
depth estimator which uses the MLE of the photon flux waveform.  Further
improvements can be obtained by incorporating spatial priors in a regularized
optimization framework \cite{Heide:2018:subpicosecond}, or data-driven neural
network-based approaches \cite{Lindell} that exploit spatial correlations
between neighboring pixels and across different training images to improve
depth accuracy.
% \smallskip

\noindent{\bf Extension to other active-imaging modalities:}
The idea of using asynchronous acquisition schemes can be extended to other SPAD-based
active-imaging applications that use the principle of TCSPC to recover the true
shape of the photon flux waveform. Non-uniform shifting schemes may be required
for time-domain FLIM where true waveform shape is an exponential decay and NLOS
imaging where the photon flux waveform can have arbitrary shapes.

{\small
\bibliographystyle{ieee_fullname}
\bibliography{sample-bibliography}
}

\clearpage
\onecolumn
\renewcommand{\figurename}{Supplementary Figure}
\renewcommand{\thesection}{Supplementary Note \arabic{section}}
\renewcommand{\theequation}{S\arabic{equation}}
\setcounter{figure}{0}
\setcounter{section}{0}
\setcounter{subsection}{0}
\setcounter{equation}{0}
\setcounter{page}{1}

\begin{center}
%\begin{tabular}{c}
\huge Supplementary Document for\\[0.2cm]
\huge ``\mytitle'' \\[1.1cm]
\normalsize Anant Gupta, Atul Ingle, Mohit Gupta.\\[0.2cm]
\texttt{\{anant,ingle,mohitg\}@cs.wisc.edu}
%\end{tabular}
\end{center}

% STOP --- READ THIS %
% Do not reorder or insert new section commands for the first 8 sections.
\section{Asynchronous Image Formation Model and MLE Waveform Estimator \label{sn:async_image_formation}}
In this supplementary note we derive the Poisson-Multinomial histogram model of
Eq.~(\ref{eq:histogram_joint_prob}) in the main paper. We then derive the generalized Coates's
estimator (Eq.~(\ref{eq:modified_coates_estimator})), which is the maximum
likelihood estimator (MLE) of the photon flux waveform in asynchronous acquisition. Scene depth is computed by
locating the peak of the estimated photon flux waveform.

\subsection*{Joint Distribution of Measured Histogram}
Here we derive the joint distribution of the histogram $(N_1,\ldots, N_B,
N_{B+1})$ measured in the asynchronous acquisition mode.  Recall that the
$B+1^\text{st}$ bin is added for mathematical convenience to record SPAD cycles
with no detected photons.  For the $l^\text{th}$ SPAD cycle we define a one-hot
random vector $(O_{l,1}, O_{l,2},\ldots,O_{l,B},O_{l,B+1})$ that stores the bin
index where the photon was recorded. Since the SPAD detects at most one photon
per laser cycle, $(O_{l,i})_{i=1}^{B+1}$ contains zeroes everywhere except at
the bin index corresponding to the photon detection.  Its joint distribution is
given by a categorical distribution [Suppl.~Ref.~\ref{ref_murphy}]:
\begin{equation}
  (O_{l,i})_{i=1}^{B+1} \sim \text{($B\!+\!1$)-Categorical}\big((p_{l,i})_{i=1}^{B+1}\big).
  \label{eq:categorical}
\end{equation}
The final histogram of photon counts is obtained by summing these one-hot
vectors over all laser cycles:
\begin{equation} 
  N_i = \sum_{l=1}^L O_{l,i} \, .
  \label{eq:hist} 
\end{equation}
Since $N_i$ is a sum of $L$ different \text{($B\!+\!1$)-Categorical} random
variables, the joint distribution is given by a Poisson-Multinomial
Distribution [Suppl. Ref. \ref{ref_pmd}]:
\begin{equation}
  \hspace{-0.1in} (N_i)_{i=1}^{B+1} \sim \text{$(L,B\!+\!1)$-PMD}\bigg( (p_{l,i})_{1\leq l\leq L, 1\leq i\leq B+1} \bigg).
  \label{eq:histogram_joint_prob}
\end{equation}
The expected number of photon counts  $\E [N_i]$ in the $i^\text{th}$ bin is
$\sum_{l=1}^L p_{l,i}$. Note that in the synchronous case, this reduces to a
multinomial distribution because the one-hot random vector defined here no
longer depends on the SPAD cycle index $l$.

%Recall that the indicator random variable $D_{l,i}$ (Def.~\ref{def:den_seq}) is
%$0$ if the SPAD has already detected a photon in the $l^\text{th}$ SPAD cycle
%in a histogram bin that precedes (in a modulo-B sense) the $i^\text{th}$
%histogram bin, thus preventing the SPAD from detecting a photon in the
%$i^\text{th}$ bin. This can be expressed in terms of $O_{l,i}$ as:
%\begin{equation} \label{denom}
%  D_{l,i} = 1 - \sum_{j \in J_{l,i}} O_{l,j}.
%\end{equation}

\subsection*{Derivation of the Generalized Coates's Estimator for Asynchronous SPAD LiDAR (Eq.~(\ref{eq:modified_coates_estimator}))}
In this section, we derive the MLE $(\widehat r_i)_{i=1}^B$ of the true
waveform $(r_i)_{i=1}^B$ for the asynchronous acquisition model and show that
it is equal to the generalized Coates's estimator described in the main text.
We assume that for each SPAD cycle $1\leq l \leq L$, the TCSPC system stores
one-hot random vectors $(O_{l,i})_{1\leq i \leq B+1}$.

For future reference, we define $J_{l,i}$ to be the set of bin indices
preceding $i$ in the $l^\text{th}$ cycle, in a modulo-B sense\footnote{For
example, suppose $B\!=\!8$ and $s_l\!=\!3$. Then, $J_{l,7}\!=\!\{4,5,6\},$
and $J_{l,2} \!=\!  \{4,5,6,7,8,1\}$.}:
\begin{equation}
J_{l,i}\! =\! 
\begin{cases}
  \{s_l\!+\!1,\ldots,\! \underbrace{B, 1}_\text{wrap around},\! \ldots, i\!-\!1\}, &\text{for } i\leq s_l \\
  \{s_l\!+\!1, \ldots, i\!-\!1\}, &\text{for } i>s_l\,. \label{eq:index_set}
\end{cases}
\end{equation}

In the $l\text{th}$ laser cycle, the joint distribution of
$(O_{l,i})_{i=1}^{B+1}$ is given by the categorical distribution in
Eq.~(\ref{eq:categorical}). Therefore, the likelihood function of the photon
incidence probabilities $(q_i)_{i=1}^B$ is given by:

\begin{align}
  \mathcal{L}(q_1, q_2, ..., q_B) &= \P\left((O_{1,i})_{i=1}^{B+1}, (O_{2,i})_{i=1}^{B+1}, ..., (O_{L,i})_{i=1}^{B+1} | q_1, q_2, ..., q_B\right) \nonumber \\
    &\overset{(a)}{=} \prod_{l=1}^L \P\left((O_{l,i})_{i=1}^{B+1} | q_1, q_2, ..., q_B\right) \nonumber \\
    &\overset{(b)}{=} \prod_{l=1}^L  \prod_{i=1}^{B+1} p_{l,i}^{O_{l,i}} \nonumber \\
    &\overset{(c)}{=} \prod_{l=1}^L \prod_{i=1}^{B+1} {\left( q_i \prod_{j \in J_{l,i}} (1 - q_j) \right)}^{O_{l,i}} \nonumber  \\
    %&= \left[ \prod_{i=1}^{B+1} \prod_{l=1}^{L} q_i^{O_{l,i}} \right] \left[ \prod_{l=1}^{L} \prod_{i=1}^{B+1}\left( \prod_{j \in J_{l,i}}  (1 - q_j) \right)^{O_{l,i}}\right]  \\
    &\overset{(d)}{=} \prod_{i=1}^{B+1} q_i^{\sum_{l=1}^L O_{l,i}} \left[ \prod_{l=1}^{L} \prod_{i=1}^{B+1} \left( \prod_{j = 1}^{B+1} (1 - q_j)^{\mathbbm{1}(j \in J_{l,i})} \right)^{O_{l,i}}\right] \nonumber  \\
    %&\overset{(e)}{=} \prod_{i=1}^{B+1} q_i^{N_i} \left[ \prod_{j = 1}^{B+1} \left( \prod_{l=1}^{L} \prod_{i=1}^{B+1}  (1 - q_j)^{\mathbbm{1}(j \in J_{l,i}) {O_{l,i}}} \right)\right]  \\
    %&\overset{(e)}{=} \prod_{i=1}^{B+1} q_i^{N_i} \prod_{j = 1}^{B+1} \prod_{l=1}^{L} \prod_{i=1}^{B+1}  (1 - q_j)^{\mathbbm{1}(j \in J_{l,i}) {O_{l,i}}} \\
    &\overset{(e)}{=} \prod_{i=1}^{B+1} q_i^{N_i} \prod_{j = 1}^{B+1} \prod_{l=1}^{L} (1 - q_j)^{\sum_{i=1}^{B+1}\mathbbm{1}(j \in J_{l,i}) {O_{l,i}}} \nonumber \\
    %&= \prod_{i=1}^{B+1} q_i^{N_i} \left[ \prod_{j = 1}^{B+1} \left( \prod_{l=1}^{L} (1 - q_j)^{\sum_{i=1}^{B+1}\mathbbm{1}(j \in J_{l,i}) {O_{l,i}}} \right)\right]  \\
    &\overset{(f)}{=} \prod_{i=1}^{B+1} q_i^{N_i} \prod_{j = 1}^{B+1} \prod_{l=1}^{L} (1 - q_j)^{D_{l,j} - O_{l,j}} \nonumber \\
    %&= \prod_{i=1}^{B+1} \left[q_i^{N_i}\right] \left[ \prod_{j = 1}^{B+1} \left( (1 - q_j)^{\sum_{l=1}^L (D_{l,j} - O_{l,j})} \right)\right]  \\
    &\overset{(g)}{=} \prod_{i=1}^{B+1} q_i^{N_i} (1 - q_i)^{D_i - N_i} \label{eq:likelihood}
    %&= \prod_{i=1}^{B+1} \left[q_i^{N_i} (1 - q_i)^{D_i - N_i} \right]  \\
\end{align}
where (a) holds because measurements in different cycles are conditionally
independent given the shift sequence; (b) follows from the definition of
categorical distribution and the fact that for any fixed $l$, $O_{l,i}=1$ for
exactly one $1\leq i\leq B+1$; (c) follows from
Eq.~(\ref{eq:photon_detection_prob}); (d) uses the notation $\mathbbm{1}$ for
the indicator function; (e) uses the definition of $N_i$ from
Eq.~(\ref{eq:hist}); (f) follows from the lemma proved below; (g) follows from
Eq.~(\ref{eq:hist}) and Def.~\ref{def:den_seq} and rearrangement of the terms in
the preceding product.

Since the likelihood is factorizable in $q_i$, we can calculate the MLE element-wise as:
\begin{align*}
  \widehat q_i &= \argmax_{q_i} \; q_i^{N_i} (1 - q_i)^{D_i - N_i} \\
    &= \frac{N_i}{D_i}.
\end{align*}
Since $q_i = 1-e^{-r_i}$, by the functional invariance property of the MLE
[Suppl. Ref. \ref{ref_kay}], the MLE for the photon flux waveform $r_i$ is
given by the generalized Coates's estimator of
Eq.~(\ref{eq:modified_coates_estimator}).  We estimate scene depth by locating
the peak of the estimated waveform: $\widehat \tau = \argmax_i \widehat r_i.$

Finally, we prove the following lemma that was used in step (f) in the derivation above.
\newtheorem*{lemma*}{Lemma}
\begin{lemma*}[\bf Proof of step (f)]
For $1 \leq j \leq B+1$ and $1 \leq l \leq L$
 \[ \sum_{i=1}^{B+1}\mathbbm{1}(j \in J_{l,i}) {O_{l,i}} = D_{l,j} - O_{l,j} \]
\end{lemma*}
\begin{proof}
  Let $i^*$ denote the bin index where the photon was detected in the
  $l^\text{th}$ SPAD cycle, i.e., $O_{l,k} = 1$ iff $k = i^*$ and
  $0$ otherwise.  Then: 
  \[ 
    \sum_{i=1}^{B+1}\mathbbm{1}(j \in J_{l,i}) {O_{l,i}} = \mathbbm{1}(j \in J_{l,i^*}) {O_{l,i^*}}.
  \] 
  If $j = i^*$, then by definition $j \not\in J_{l, i^*}$.  Therefore, LHS = 0.
  Also, $D_{l,j} = O_{l,j} = 1$ in this case, so RHS = 0.  If $j \neq i^*$,
  there are two cases.  Case 1: $j \in J_{l, i^*}$. Then $D_{l,j} = 1$ and
  $O_{l,j} = 0$.  Therefore, LHS = RHS = 1. Case 2: $j \not\in J_{l, i^*}$.
  Then $D_{l,j} = O_{l,j} = 0$.  Therefore, LHS = RHS = 0.
\end{proof}

\clearpage\newpage
\section{Proofs of Results~\ref{thm1} and \ref{thm2}\label{sn:res1_res2_proofs}}
In this section we provide detailed mathematical proofs for two key theoretical
results in the main paper. Recall that we use an upper bound on the probability
of depth error ($\ell_0$ error) as a surrogate for RMSE. Result~\ref{thm1}
establishes the importance of a constant expected denominator sequence. It
shows that a shifting strategy that minimizes an upper bound on the $\ell_0$
error must have a denominator sequence that is constant (on average) for all
histogram bins. Result~\ref{thm2} shows that the uniform shifting strategy
(which allocates approximately equal number of shifts to all histogram bins
over whole depth range) achieves a constant expected denominator sequence.

\subsection*{Proof of Result 1}
In this section, we will derive an upper bound on $\ell_0$ depth error which
corresponds to the probability that the depth estimate using the generalized
Coates's estimator derived in \ref{sn:async_image_formation} is different from
the true depth bin.

\paragraph{An upper bound on \bm{$\ell_0$} error:} To ensure that the depth
estimate $\widehat \tau$ is correct, the bin corresponding to the true depth
should have the highest counts after the generalized Coates's correction
(Eq.~(\ref{eq:modified_coates_estimator})) is applied. Therefore, for a given
true depth $\tau$, we want to minimize the probability of depth error:
\begin{align*}
  \P(\widehat \tau \neq \tau) &= \P\left(\bigcup_{i\neq \tau} (\widehat q_i > \widehat q_\tau)\right) \\
     & \leq \sum_{i \neq \tau} \P\left(\widehat q_i > \widehat q_{\tau}\right) \\
     &= \sum_{i \neq \tau} \P\left(\widehat q_i - \widehat q_{\tau} > 0 \right)
\end{align*}
where the first inequality follows from the union bound. Note that $\widehat
q_i - \widehat q_\tau$ has a mean $q_i - q_\tau$ and variance $\sigma^2_i +
\sigma^2_\tau$ (assuming uncorrelated). The variance is given by [Suppl.  Ref.~\ref{ref_gupta}] $\sigma_i^2 =
\frac{q_i (1-q_i)}{\E[D_i]}.$

For large $L$, by the central limit theorem, we have:
\[ \widehat q_i - \widehat q_\tau \sim \mathcal{N}(q_i - q_\tau, \sigma^2_i + \sigma^2_\tau). \]
%\[ = \sum_{i \neq \tau} \P\left(\widehat p(i) > \widehat p(\tau)\right) \]
Using the Chernoff bound for Gaussian random variables, we get:
\begin{align*}
 \P(\widehat q_i > \widehat q_\tau) &\leq \exp{\left( - \frac{(q_i - q_\tau)^2}{2 (\sigma^2_i + \sigma^2_\tau)} \right)} \\
 &= \frac{1}{2}\exp{\left( - \frac{1}{2}\frac{(q_i - q_\tau)^2}{ \frac{q_i (1 - q_i)}{\E[D_i]} + \frac{q_{\tau} (1 - q_{\tau})}{\E[D_{\tau}]}} \right)}
 \end{align*}
Assuming a uniform prior on $\tau$ over the entire depth range, we get the
following upper bound on the average probability of error:
\begin{align}
\label{eq:upperbound}
 \frac{1}{B} \sum_{\tau=1}^B \P(\widehat \tau \neq \tau)  &\leq \frac{1}{B}\sum_{\tau=1}^B \sum_{i \neq \tau} \frac{1}{2}\exp{\left( - \frac{1}{2}\frac{(q_i - q_\tau)^2}{ \frac{q_i (1 - q_i)}{\E[D_i]} + \frac{q_{\tau} (1 - q_{\tau})}{\E[D_{\tau}]}} \right)} \nonumber \\
 &\approx \frac{1}{B}\sum_{i,\tau=1}^B \frac{1}{2}\exp{\left( - \frac{1}{2}\frac{(q_i - q_\tau)^2}{ \frac{q_i (1 - q_i)}{\E[D_i]} + \frac{q_{\tau} (1 - q_{\tau})}{\E[D_{\tau}]}} \right)}
\end{align}

We can minimize the probability of error indirectly by minimizing this upper
bound. The upper bound involves exponential quantities which will be dominated
by the least negative exponent, which in turn is dominated by the index $i$
with the largest value of $\nicefrac{1}{\E[D_i]}.$ Therefore, the denominator
sequence that minimizes this upper bound must maximize $\min_i \E[D_i]$.
Given that the total expected denominator is constant under a fixed number of cycles (see proof below), this
is equivalent to making the denominator sequence uniform.

The above analysis assumes that photon detections across cycles are independent, and therefore holds for all acquisition schemes with a gating mechanism that is fixed in advance, including synchronous and asynchronous
acquisition (deterministic). Note that it does not hold for photon-driven shifting, where a photon detection in one cycle can affect that in another.
% Note that this analysis did not make any assumptions about the acquisition
% scheme used and holds for a general expected denominator sequence. It holds for
% all TCSPC-based acquisition schemes, including synchronous and asynchronous
% acquisition (both deterministic and photon driven).

\paragraph{Proof of constant total expected denominator:}
% We assume a nearly uniform rate function with $r_i \approx \lam$ for all $i$. In any SPAD cycle $l$, the denominator sum $\sum_{i=1}^B D_{l,i} = A_l$, where $A_l$ denotes the arrival time of the first photon, measured from the start of the SPAD cycle. If no photon is detected, $A_l = m$. Therefore, the total expected denominator is given by:
% \begin{align*}
% \Xi &\equiv \E[\sum_{i=1}^B D_{i}] \\
% &= \E[\sum_{l=1}^L \sum_{i=1}^B D_{l, i}] \\
% &= \sum_{l=1}^L \E[A_l] \\
% \end{align*}
% Now:
% \begin{align*}
%     \E[A_l] &= \sum_{i=1}^m \P(A_l = i) \cdot i + \left(1 - \sum_{i=1}^m \P(A_l = i) \right) \cdot m \\ 
%     &= \sum_{i=1}^m e^{}
% \end{align*}

Let $\Xi \equiv \sum_{i=1}^B \E[D_i]$ be the total expected denominator.
% For our analysis, we only care about the average photon flux.
Assuming a low SBR scenario where the background flux dominates the signal
flux, we take $r_i \approx \lam$ for all $i$, i.e., an almost uniform incident
waveform. To calculate the total denominator, we sum up the contributions of
each cycle. If the SPAD active time is $m$, each cycle ends with either a
photon detection in one of the $m$ time bins, or with no photon
detections. A cycle with a photon detection in the $i^\text{th}$ bin ($1 \leq i
\leq m$) contributes $i$ units to the total denominator, since each bin before
and including the detection bin was active. By the same argument, a cycle with
no photon detections contributes $m$ units. Therefore, the total expected
denominator is given by:
\begin{align}
    \Xi &= \sum_{l=1}^L \left[ \sum_{i=1}^m \P(O_{l,s_l \oplus i} = 1) \cdot i + \P(O_{l,s_l \oplus i} = 0 \text{ for } 1 \leq i \leq m) \cdot m \right]\nonumber \\
    &= \sum_{l=1}^L \left[ \sum_{i=1}^m p_{l, s_l \oplus i} \cdot i + \left(1 - \sum_{j=1}^m p_{l, s_l \oplus j}\right) \cdot m \nonumber \right] \\
    &\overset{(a)}{=} \sum_{l=1}^L \left[ \sum_{i=1}^m (1 - e^{-\lam}) e^{- (i-1) \lam} \cdot i + e^{-m \lam} \cdot m \right] \nonumber \\
    % &= 1 + e^{-\lambda} + e^{-2 \lambda} + ... + e^{-(m-1) \lambda} \nonumber \\
    &= \frac{L(1 - e^{-m \lam})}{1 - e^{-\lam}} \;,
    \label{eq:D_per_N}
\end{align}
where $\oplus$ denotes addition modulo-B, and (a) follows from the fact that
for a uniform waveform, $p_{l, s_l \oplus i} = (1 - e^{-\lam})
(e^{-\lam})^{|J_{l,s_l \oplus i}|}$. Moreover, $J_{l,s_l \oplus i} = i-1$.
Therefore, $\Xi$ remains constant for a given $L$, regardless of the shift
sequence used. 

%The value of $D_i$ that maximizes the least
%negative exponent must satisfy:
%\begin{align*}
% D^{\text{opt}} &= \argmin_{D} \frac{1}{B}\sum_{i,\tau=1}^B \frac{1}{2}\exp{\left( - \frac{1}{2}\frac{(q_i - q_\tau)^2}{ \frac{q_i (1 - q_i)}{\E[D_i]} + \frac{q_{\tau} (1 - q_{\tau})}{\E[D_{\tau}]}} \right)} \\
% &\approx \argmin_{D} \max_{i, \tau} \left[ \frac{1}{2}\exp{\left( - \frac{1}{2}\frac{(q_i - q_\tau)^2}{ \frac{q_i (1 - q_i)}{\E[D_i]} + \frac{q_{\tau} (1 - q_{\tau})}{\E[D_{\tau}]}} \right)} \right] \\
% &= \argmin_{D} \max_{i} \frac{1}{\E[D_i]} \\
% &= \argmax_{D} \min_{i} \E[D_i]
%\end{align*}
%The second last step is true since the term inside the exponent is maximized
%for $i = \tau = \argmin_i \E[D_i]$. Furthermore, the expression depends
%inversely on $\E[D_i]$ and $\E[D_\tau]$, and all other quantities ($q_i, q_\tau$)
%are independent of $D$. Therefore, minimizing the expression is equivalent to
%maximizing the minimum expected denominator.

\subsection*{Proof of Result 2}
For $1 \leq i \leq B$, under uniform shifting:
\begin{align*}
  \E[D_i] &\overset{(a)}{=} \sum_{l=1}^L \E[D_{l,i}] \\
    &= \sum_{l=1}^L \prod_{j \in J_{l,i}} (1 - q_j) \\
    &\overset{(b)}{\approx} \sum_{l=1}^L \prod_{j \in J_{l,i}} e^{-\lam} \\
    &= \sum_{l=1}^L e^{-\lam |J_{l,i}|} \\
    &= \sum_{l=1}^L e^{-\lam \{(i - s_l)\bmod B\}} \\
    &\overset{(c)}{=} \frac{L}{B} \left( \sum_{i=0}^{B-1} e^{-i \lam} \right) \\
\end{align*}
where (a) follows from Def.~\ref{def:den_seq}; (b) relies on the assumption
that the pileup is mainly due to ambient light $\lam$ and not the source light $\muu$;
and (c) assumes that $L$ is a multiple of $B$. The last step follows because
under uniform shifting, the set of shifts $\{s_l\}_{l=1}^L$ spans the discrete
space $\{0, 1, ..., B-1\}$ uniformly.  Moreover, the shifts remain uniform when
transformed by a constant modulo addition. Therefore, $\E[D_i]$ is the sum of
identical terms, and hence same for all $i$.

If $L$ is not a multiple of $B$, the shift sequence cannot achieve every
possible shift from $0$ to $B-1$ an equal number of times and some bins might
use more shifts than others, especially when $L < B$.However, the expected
denominator sequence is still approximately uniform, with the approximation
becoming more accurate as $L$ increases. This can be intuitively seen from the
gradual elimination of the ``saw teeth'' in
Fig.~\ref{fig:deterministic_shifting_intuition} as the number of shifts
increases. 

% Some   approximately Since $s_l$ forms a discretized uniform partition of
% $[0, B]$, $\bmod(i - s_l, B)$ also forms a discretized uniform partition for
% all $i$.  \textcolor{red}{Assume $m=B$.  Assume $L$ is a multiple of $B$.
% Otherwise this is only approximately true. Point to Fig.~2 and talk about
% teeth.}

\clearpage\newpage
\section{Achieving Uniform Shifts through Laser and SPAD Cycle Mismatch \label{sn:fractional_shifts}}
% \subsection{Uniform Shifting via SPAD Cycle Extension \label{sec:uniform_extension}} 
The implementation shown in Fig.~\ref{fig:deterministic_shifting_intuition}
relies on the SPAD cycle period being different from the laser cycle.
By default, the duration for which the SPAD gate is kept on (called
the \emph{active time window}, shown as white boxes in
Fig.~\ref{fig:deterministic_shifting_intuition}), is set equal to the laser
cycle period $B\Delta$. After each active time window, we force the SPAD gate
to be turned off (gray boxes in
Fig.~\ref{fig:deterministic_shifting_intuition}) for a duration equal to
$\td$ (irrespective of whether a photon was detected). As a result, the SPAD
cycle length is equal to $B\Delta+\td$.  This ensures that the dead time from
one SPAD cycle does not extend into the next when a photon is detected close to
the end of the current cycle.  The total number of SPAD cycles over a fixed
acquisition time, $T$, is therefore limited to $L = \lfloor
\nicefrac{T}{(B\Delta+\td)} \rfloor$.

Since the length of the laser cycle is $B\Delta$ and the SPAD cycle is $B\Delta
+ \td$, we automatically achieve the shift sequence $(0, \td, 2\td,\ldots)$.
Moreover, we can achieve any arbitrary shift sequence in practice by
introducing additional artificial delays $(\varepsilon_l)_{l=1}^L$ (using, say, a programmable delayer) at the end
of each SPAD cycle that extend the inactive times beyond $\td$. In general, the
additional delay $\sum_l \varepsilon_l$ will require either increasing $T$ or
decreasing $L$, both of which are undesirable.  In the next section, we show that it is possible to choose
$\varepsilon_l$'s such that the total extra delay $\sum_l\varepsilon_l \leq 1.5
B\Delta$, by making the SPAD cycle length co-prime with the laser. Therefore,
uniform shifting can be implemented in practice at a negligible additional cost
in terms of total acquisition time.

\subsection*{Quantifying acquisition time cost of uniform shifting}
In this section, we bound the additional delay incurred due to uniform shifting. We consider the general case where the SPAD active time window can be different from the laser cycle period. We also present a practical method for implementing uniform shifting that relies on artificially introducing a mismatch in the laser and SPAD repetition
frequencies.
% Uniform shifting can be achieved with a n\"aive method that sequentially cycles
% through each individual shift (using, say, a programmable delayer) between the
% laser and the SPAD cycles. However, this would lead to additional delays which
% would be inefficient in practice when operating under an acquisition time
% budget. In this section we present a practical method for achieving a uniform
% shift sequence under a total acquisition time constraint. Our method relies on
% artificially introducing a mismatch in the laser and SPAD repetition
% frequencies.

% It is possible to achieve all shift amounts in the set
% $\{0,\ldots, B-1\}$ uniformly by introducing a slight mismatch between the time
% periods of the laser cycles and the SPAD measurement cycles. We artificially
% extend the dead time of the SPAD by a small amount, say, $\epsilon$.

Let $\td$ be the dead time of the SPAD, $T$ be the total acquisition time, $B$
be the number of bins in the histogram and $\Delta$ be the size of each
histogram bin. Let $N = \lfloor \nicefrac{T}{\Delta} \rfloor$ be the total
acquisition time and $n_d = \lfloor \nicefrac{\td}{\Delta} \rfloor$ be the dead
time in units of the histogram bin size. Suppose the SPAD is kept active for
$m$ bins in each cycle. Our goal is to find a small positive shift $\epsilon$
such that $m + n_d + \epsilon$ becomes asynchronous to $B$, in the sense that
uniform shifts modulo-B are achieved in $N$ bins, and in equal proportions for
all shifts.

Let $L = \lfloor \nicefrac{N}{m + n_d} \rfloor$ be the number of whole cycles that
would be obtained if no shifts were used. For simplicity, assume $B$ is a
multiple of $L$.  We divide the range $B$ into $L$ equally spaced intervals and
design a shift sequence $(\epsilon_i)_{i=1}^L$ such that each of the $L$ cycles
is aligned with the start of exactly one of these intervals.

When $\epsilon=0$, the dead time provides a shift amount equal to $\nicefrac{(m +
n_d) L}{B} \bmod L$, in units of the interval size $\nicefrac{B}{L}$. By rounding
it up to the nearest integer, we get the effective shift $ s = \nicefrac{(m +
n_d + \epsilon)L}{B} = \left\lceil \nicefrac{(m + n_d)
L}{B}\right\rceil$. We consider two cases:
\smallskip

\noindent{\bf Case 1:} Suppose $s$ and $L$ are co-prime. In this case, the $i^\text{th}$
cycle has a unique shift $(i - 1) s \bmod L$ for $1 \leq i \leq L$.
\medskip

\noindent{\bf Case 2:} Suppose $s$ and $L$ are not co-prime and their greatest
common divisor is $g$. We can express $s = g \cdot k$ and $L = g \cdot l$ for
some co-prime integers $k$ and $l$. At the end of every $l$ SPAD cycles, each
of the $l$ shifts $\{g,2g, 3g, \ldots, ...,l \cdot g\}$ is attained once. Since
we have $g$ of these ``groups'' of $l$ SPAD cycles, we can add an offset of $i$
to the shift amounts of the $i^\text{th}$ group for $i=0,1,...,g-1$. This will
ensure a unique shift for each of the $L$ cycles. 
\medskip

\noindent{\bf Additional delay due to shifting:} The additional delay due to
shifting is $L \epsilon$ in the Case 1 and $L \epsilon + (g - 1)
\nicefrac{B}{L}$ in Case 2. Since $\nicefrac{L \epsilon}{B} < 1$ and $g \leq
L/2$, the total additional delay incurred due to shifting time is negligible
and is at most $1.5 B$ bins, i.e., at most one-and-a-half extra laser cycle
periods.

\medskip
\noindent{\bf Achieving uniform shift sequence through frequency mismatch:}
Note that in both these cases, uniform shifting can be achieved conveniently by
operating the SPAD cycle frequency asynchronous to the laser frequency:
$\nicefrac{L}{B\epsilon(L + s)}$ in the Case~1 and $\nicefrac{L}{B\epsilon(L
  +s+\nicefrac{1}{l})}$ in Case~2.

% Note that if number of cycles is not a
% multiple of $B$, then the denominator sequence is only approximately constant

\clearpage\newpage
\section{Derivation of $\bm{m}^\text{opt}$ \label{sn:mopt_derivation}}
%TODO plot m_opt for a few different phi_bkg values - just for some intuition.
% \textcolor{red}{Move this part after the derivation of total expected denominator.
% Say that we have already done uniform shifts to make denominatory sequence constant on
% average. so ok to max total.}

\begin{figure}[!ht]
  \centering\includegraphics[width=0.5\columnwidth]{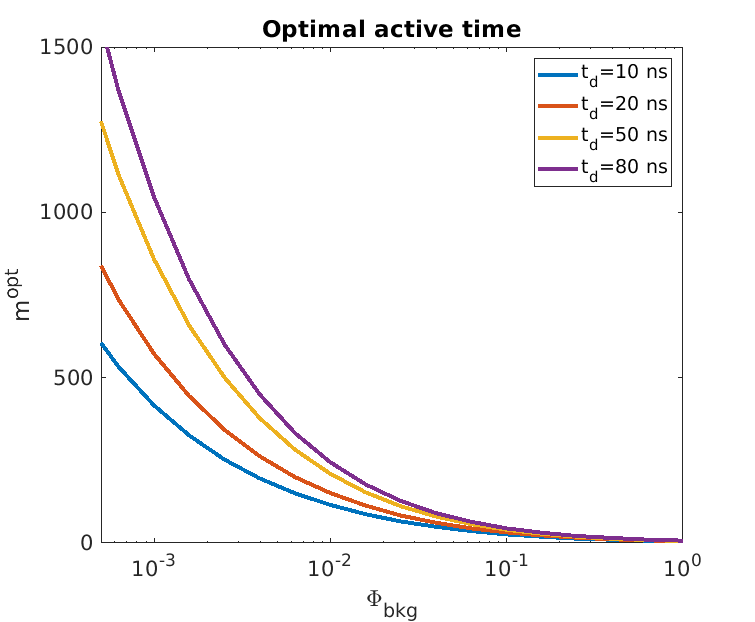}
  \caption{ {\bf Effect of ambient flux and dead time on $\bm{m}^\text{opt}$ .} The
  optimal active time is a decreasing function of ambient flux $\lam$ and an
  increasing function of dead time $\td$.
\label{fig:optimal_active_time}}
\end{figure}

In this section, we derive the optimal active time for uniform shifting. We
assume that the expected denominator sequence is constant (achieved, say, using
the method of laser and SPAD cycle mismatch from the previous section). Recall
that we use an upper bound on the $\ell_0$ depth error as a surrogate for RMSE.
From the proof of Result~\ref{thm1} we know that this is equivalent to using
the smallest denominator sequence value as a surrogate for RMSE. However, since
the denominator sequence is constant on average, we can use the total expected
denominator $\Xi$ from Eq.~(\ref{eq:D_per_N}) as a surrogate for depth
accuracy---a lower total expected denominator would correspond to higher depth
errors and vice versa.  We now derive the value of $m$ that maximizes $\Xi$.

% For our analysis, we only care about the average photon flux.
% Assuming a low SBR scenario where the background flux dominates the signal
% flux, we take $r_i \approx \lam$ for all $i$, i.e., an almost uniform
% incident waveform. To calculate the total denominator as a function of $m$, we
% sum up the contributions of each cycle. Since the SPAD active time is $m$, each
% cycle ends with either a photon detection in one of the $m$ discrete time bins,
% or with no photon detection. A cycle with a photon detection in the $i$th bin,
% $1 \leq i \leq m$, contributes $i$ to the total number of denominators, since each bin before and including the detection bin was
% active. By the same argument, a cycle with no photon detections contributes $m$. Therefore, the total expected denominator is given by:
% \begin{align}
%     \Xi(m) &= \sum_{l=1}^L \left[ \sum_{i=1}^m \P(O_{l,s_l \oplus i} = 1) \cdot i + \P(O_{l,s_l \oplus i} = 0 \text{ for } 1 \leq i \leq m) \cdot m \right]\nonumber \\
%     &= \sum_{l=1}^L \left[ \sum_{i=1}^m p_{l, s_l \oplus i} \cdot i + \left(1 - \sum_{j=1}^m p_{l, s_l \oplus j}\right) \cdot m \nonumber \right] \\
%     &= \sum_{l=1}^L \left[ \sum_{i=1}^m (1 - e^{-\lam}) e^{- (i-1) \lam} i + e^{-m \lam} m \right] \nonumber \\
%     % &= 1 + e^{-\lambda} + e^{-2 \lambda} + ... + e^{-(m-1) \lambda} \nonumber \\
%     &= \frac{L(1 - e^{-m \lam})}{1 - e^{-\lam}} \;,
%     \label{D_per_N}
% \end{align}
% where $\oplus$ denotes addition modulo B.\\
% where $u$ denotes the photon detection bin when $1 \leq u \leq m$, and no photon detection when $u = m+1$. \\
Since the length of each SPAD cycle is $m\Delta + \td$, the number of cycles in
a fixed acquisition time $T$ is given by
\begin{align}
    L = \frac{T}{m\Delta + \td}.
    \label{eq:N}
\end{align}
From Eqs.~(\ref{eq:D_per_N}) and (\ref{eq:N}), we get:
% \begin{align*}
%     \Xi &= \frac{T(1 - e^{-m \lam})}{(m\Delta + \tau)(1 - e^{-\lam})}
% \end{align*}
%  \[ \frac{T}{m+\tau} (n\mod{m + \tau}) = k n \]
%  and
%  \[ \gcd{n, m+\tau} = 1 \]
%  Assuming $\mu << n \lambda$, we have:
%  \begin{align*}
%  D_i &= \frac{T}{n(m + \tau)} \left(1 + e^{-\lambda} + e^{-2 \lambda} + ... + e^{-(m-1) \lambda} \right) \\
%  &= \frac{T (1 - \exp{(-m \lambda)})}{n(m + \tau)(1 - \exp{(-\lambda)})}
%  \end{align*}
%  for all $1 \leq i \leq n$.
% The expression for $\Xi$ depends on $m$ through two opposing terms
% that drive it to zero at either extreme. Clearly, the optima should lie
% somewhere in between. Treating $m$ as a continuous variable, the optimal
% solution can be found by solving the equation:
% \[ \frac{d \Xi}{d m} = 0 \]
% \[ \lambda e^{- m \lambda} (m + \tau) = 1 - e^{- m \lambda} \]
% \[ m^{\text{opt}} = -\frac{W(-e^{-\tau \lambda - 1}) + \tau \lambda + 1}{\lambda} \]
% where $W$ is the Lambert W function \cite{LambertW_1996}.

% The total expected
% denominator $\Xi = \sum_{i} \E[D_i]$ is given by:
\[
  \Xi (m) = \frac{T}{m\Delta+\td}\frac{1-e^{-m \lam}}{1-e^{-\lam}}.
\]
where we have explicitly included the dependence of $\Xi$ on the active time $m$.
Intuitively, the first term represents the average number of SPAD measurement
cycles that can fit in time $T$ and the second term is the expected denominator
value in each cycle (assumed to be uniform over all histogram bins in each
cycle). The optimal active time $m^\text{opt}$ is the one that maximizes the
total denominator. Solving for $\nicefrac{d\Xi}{dm}=0$, yields
[Suppl.~Ref.~\ref{ref_lambertw}]:
\begin{equation}
    m^\text{opt} = -\frac{1}{\Phi_\text{bkg}} \mathsf{LambertW}(-e^{-t_\text{d} \Phi_\text{bkg} / \Delta - 1}) - \frac{t_\text{d}}{\Delta} - \frac{1}{\Phi_\text{bkg}}.
  \label{s:eq:optimal_m}
\end{equation}

\noindent{\bf Interpreting $\bm{m}^\text{opt}$:}
% (Eq.~(\ref{eq:optimal_m}) in main text)
% $m^{\text{opt}}$ decreases as ambient flux $\lam$ increases.  Under high
% ambient photon flux, the average time until the first photon detection after
% the SPAD active time window begins is small. As a result, keeping the SPAD
% active for a longer duration is inefficient because it unnecessarily increases
% the length of the measurement cycle, and reduces the number of measurements
% that can be acquired over a fixed acquisition time.  Conversely, at lower flux
% levels, the optimal active time increases to match the increased average
% inter-photon arrival time.
% \smallskip
Suppl. Fig.~\ref{fig:optimal_active_time} shows the behavior of $m^\text{opt}$
for different values of dead time $t_\text{d}$ and varying ambient flux $\Phi_\text{bkg}$. Observe that the
optimal active time decreases with increasing $\lam$.
Under high
ambient photon flux, the average time until the first photon detection after
the SPAD active time window begins is small. As a result, keeping the SPAD
active for a longer duration is inefficient because it unnecessarily increases
the length of the measurement cycle, and reduces the number of measurements
that can be acquired over a fixed acquisition time.  Conversely, at lower flux
levels, the optimal active time increases to match the increased average
inter-photon arrival time.

For a fixed background flux level, $m^\text{opt}$ is higher for longer dead
times. When the dead time is long, increasing the active time duration increases
the probability of detecting a photon while not increasing the measurement cycle length
by much. 

A uniform shift sequence with SPAD active time equal to $m^\text{opt}$ can be
implemented using the method described in \ref{sn:fractional_shifts}.

\clearpage\newpage
\section{Photon-Driven Shifting: Histogramming and Error Analysis \label{sn:photon_driven_denominator}}
In this section we provide a proof of Result~\ref{thm3} which states that
photon-driven shifting achieves a uniform shift sequence for sufficiently large
acquisition times.

This section also proposes an algorithm for computing the generalized Coates's
estimator for photon-driven shifting. Unlike uniform shifting where the shift
sequence is deterministic and can be pre-computed, the shift sequence in
photon-driven shifting is random and depends on the actual photon detection
times. We show how the arrival timestamps can be used to compute $D_i$ and
$N_i$ needed for the generalized Coates's estimator. It is also possible to
estimate the waveform in free-running acquisition using a Markov chain-based
model of the arrival times~[Suppl.~Ref.~\ref{ref_rapp}]. While the Markov
chain-based estimator is based on solving a non-convex optimization problem,
the proposed generalized Coates's estimator has a closed-form expression. Both
achieve equivalent performance in terms of depth recovery accuracy. 

\subsection*{Proof of Result \ref{thm3}}
Let $(S_l)_{l=1}^L$ denote the stochastic shift sequence, where $0 \leq S_l
\leq B-1$ and $\ominus$ denote subtraction modulo-B with a wrap around when the
index falls below $1$. This shift sequence forms a Markov chain with state
space $[0, B-1]$ and transition density given by [Suppl.~Ref.~\ref{ref_rapp}]:
\begin{align*}
f_{S_{i+1} | S_i}(s_{i+1}|s_i) =
\begin{cases}
     \left(\frac{1 - e^{-\lam}}{1 - e^{- B \lam}} \right) e^{-(s_{i+1} \ominus s_i + B - t_d) \lam} &\text{if } s_{i+1} \ominus s_i < \td \\
     \left(\frac{1 - e^{-\lam}}{1 - e^{- B \lam}} \right) e^{-(s_{i+1} \ominus s_i - t_d) \lam} &\text{otherwise.}
\end{cases}
\end{align*}
The uniform distribution $f(s) = \frac{1}{B}$ is a stationary distribution for
this irreducible aperiodic Markov chain, and hence it converges to its
stationary distribution [Suppl.~Ref.~\ref{ref_grimmett}] as
$L\rightarrow\infty$. Let $f^k(s)$ denote the distribution of the $k^\text{th}$
shift. We have:

\[
  \lim_{k \to \infty} f^k(s) = f(s)
\]

Therefore, as $L \rightarrow \infty$, the empirical distribution of $(s_1, s_2,
\ldots, s_L) \to f(s)$ and all shifts are achieved with equal probability
making the shift sequence uniform. This also leads to a constant expected
denominator sequence as shown in the simulations below.
\medskip

\begin{figure}[!ht]
  \centering
   \includegraphics[width=1.0\columnwidth]{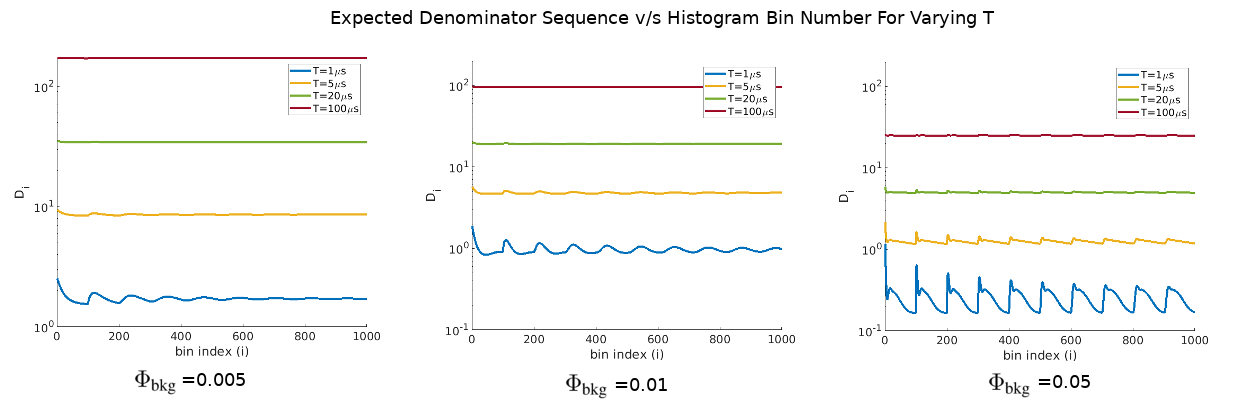}
  \caption{ {\bf Effect of flux and acquisition time on denominator sequence.} 
  The expected denominator sequence in the photon-driven mode has a position
  dependent bias which disappears as the total acquisition time increases.
\label{fig:denominator_bias}}
\end{figure}

\noindent{\bf Simulations:} Suppl. Fig.~\ref{fig:denominator_bias} shows the expected
denominator sequence for different total acquisition times at three different
ambient flux levels. There are two main observations here. First, for short
acquisition times there is a depth-dependent bias which disappears as $T$
increases.  Second, for fixed $T$, the depth dependent bias is higher for
higher flux levels.  This is because at high ambient light, the SPAD detects a
photon almost deterministically after each dead time window elapses which
causes the Markov chain $(S_l)_{l=1}^L$ to have a longer mixing time than at
lower ambient flux levels.

% Moreover,
% the variance also decreases with increasing $T$.

\subsection*{Histogram and denominator sequence computation}
In this section, we provide details about the algorithm for computing $N_i$ and
$D_i$ from the shift sequence and the sequence of photon arrival times. This
leads to a computationally tractable method for computing the generalized Coates's 
estimate for the flux waveform and hence estimating scene depths.

Let $(u_1, u_2, \ldots, u_L)$ denote the photon arrival times (in terms of bin
index) in each SPAD cycle measured with respect to the most recent laser cycle.
% Note that some SPAD cycles may have no photon detections.  We let $u_i = B+1$
% for such cycles.
Note that $1 \leq u_i \leq B$.

% To make a histogram of photon counts in each depth bin, the photon arrival
% times $u_i$ should be computed with respect to the depth origin, i.e., the
% start of the laser cycles.

The histogram of photon counts is given by:
\[
  N_i = \sum_{l=1}^L \mathbbm{1}(u_l = i)
\]
To compute the denominator sequence, we loop through the list of photon arrival
times $u_l$. For each photon detection time we increment $D_j$ for every bin
index $j\in J_{l,u_l}$.

In the photon-driven shifting mode, the denominator sequence can also be
computed in closed form. The shift sequence $(s_l)_{l=1}^L$ is determined by
the photon arrival times $u_i$ as $s_{l+1} = u_l \oplus n_d$ where $n_d$ is the
dead time in units of bins. As before, $N_i$ is given by the histogram of
$(u_l)_{l=1}^L$. $D_i$ can be computed in closed form in terms of the
histogram counts:
For each bin index $i$, there are
$\frac{T}{B\Delta}$ depth bins in total which can potentially detect photons.
However, a photon detection in any depth bin prohibits the $n_d$ bins
that follow it from detecting photons. Therefore, the value of $D_i$ at the
$i^\text{th}$ bin is given by subtracting these bins from the total:
\[
  D_i = \frac{T}{B\Delta} - \sum_{j=1}^{t_d/\Delta} N_{i\ominus j}.
\]

As before for the case of deterministic shifting, the likelihood function of the photon
incidence probabilities $(q_i)_{i=1}^B$ is given by: 
\begin{align*}
  \mathcal{L}(q_1, q_2, ..., q_B) =
    %&= \prod_{i=1}^{B+1} \left[q_i^{N_i}\right] \left[ \prod_{j = 1}^{B+1} \left( (1 - q_j)^{\sum_{l=1}^L (D_{l,j} - O_{l,j})} \right)\right]  \\
     \prod_{i=1}^{B+1} q_i^{N_i} (1 - q_i)^{D_i - N_i} \,.
    %&= \prod_{i=1}^{B+1} \left[q_i^{N_i} (1 - q_i)^{D_i - N_i} \right]  \\
\end{align*}
Therefore, the generalized Coates's estimator for photon-driven shifting is given as the following closed-form expression:
\begin{align}
\widehat \tau &= \argmax_i \frac{N_i}{D_i} \nonumber \\
&= \argmax_i \frac{N_i}{\frac{T}{B\Delta} - \sum_{j=1}^{t_d/\Delta} N_{i\ominus j}} \label{eq:photon_driven_coates} \,.
\end{align}

\subsection*{Markov chain model-based estimator [Suppl. Ref. \ref{ref_rapp}]}
\begin{figure}[!ht]
  \centering\includegraphics[width=0.3\columnwidth]{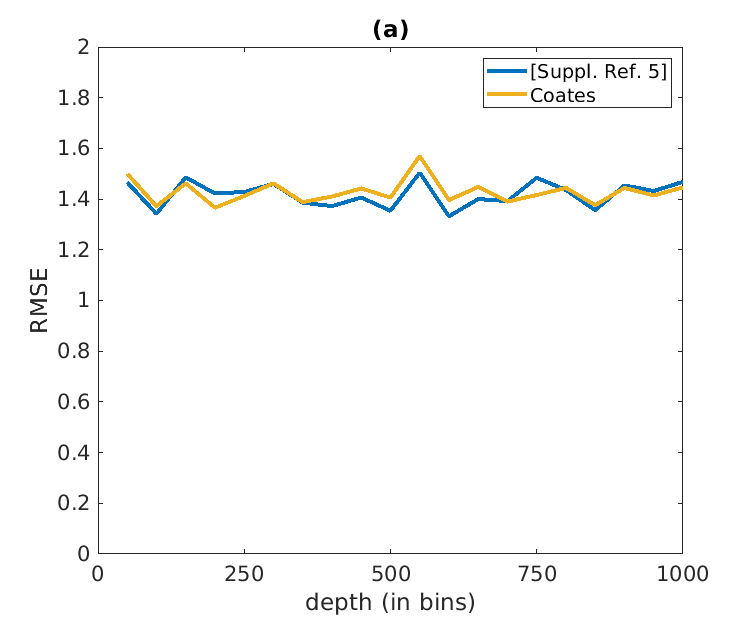}
  \includegraphics[width=0.3\columnwidth]{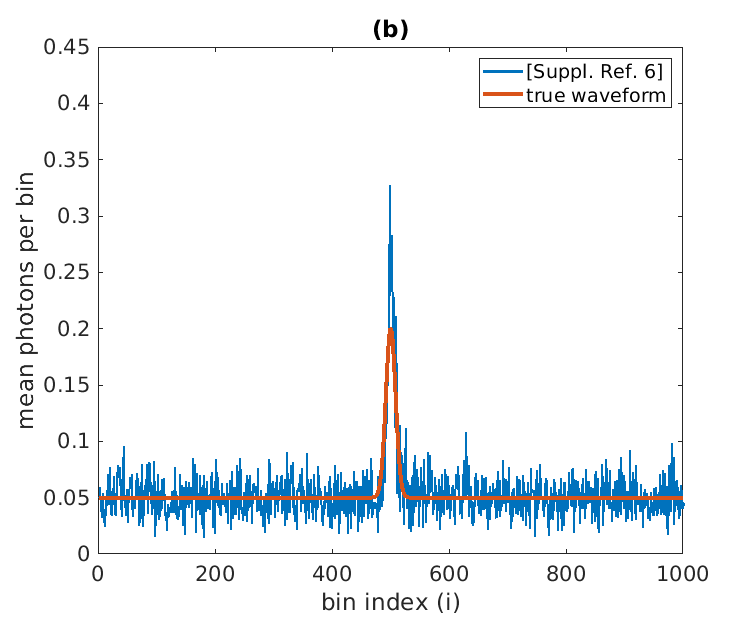}
  \includegraphics[width=0.3\columnwidth]{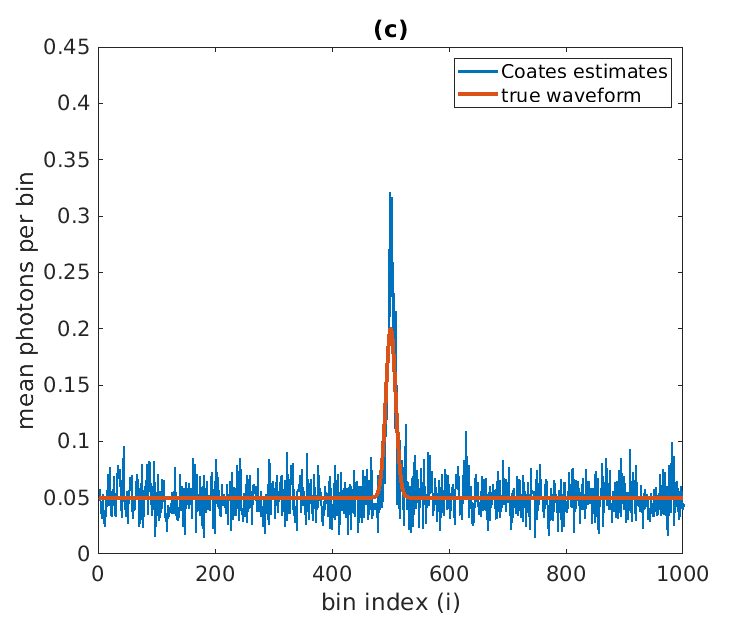}
  \caption{ {\bf Comparison between [Suppl. Ref. \ref{ref_rapp}] and the proposed generalized Coates's estimator.}
  (a) shows the RMSE curve for the two methods. Both methods have the same
  performance across depth. (b) and (c) show that the estimated waveforms using the
  two methods are similar. 
  \label{fig:rapp_and_goyal}}
\end{figure}

It is also possible to use a Markov chain for modeling the photon arrival times in free-running acquisition. The resulting estimator [Suppl. Ref. \ref{ref_rapp}] is based on solving a non-convex optimization problem, whereas the proposed generalized Coates's estimator has a closed-form expression, as shown above. Both achieve equivalent performance in terms of depth recovery accuracy, as shown in Suppl. Fig.~\ref{fig:rapp_and_goyal}. In the regime of large acquisition time ($T = \SI{25}{\micro\second}$), and with $\lam = 0.05, \muu=0.15$, the performance of both the estimators in terms of depth error is almost equivalent. 

%Crucially, however, their algorithm does not use knowledge of the total acquisition time $T$, even though it is typically known. This makes the MLE computation intractable and requires solving a non-convex optimization problem. 

%This suggests that the parameter $T$ does not carry any information about the depth that is not already contained in the histogram. However, from a computational perspective, knowledge of $T$ leads to a computationally tractable closed from depth estimator. Moreover, it also allows us to cast photon-driven shifting in the framework of asynchronous acquisition techniques, analyze its performance, and theoretically analyze its combination with optimal attenuation methods.

\subsection*{Performance Gain as a Function of Depth}

\begin{figure}[!ht]
  \centering
  \includegraphics[width=0.6\textwidth]{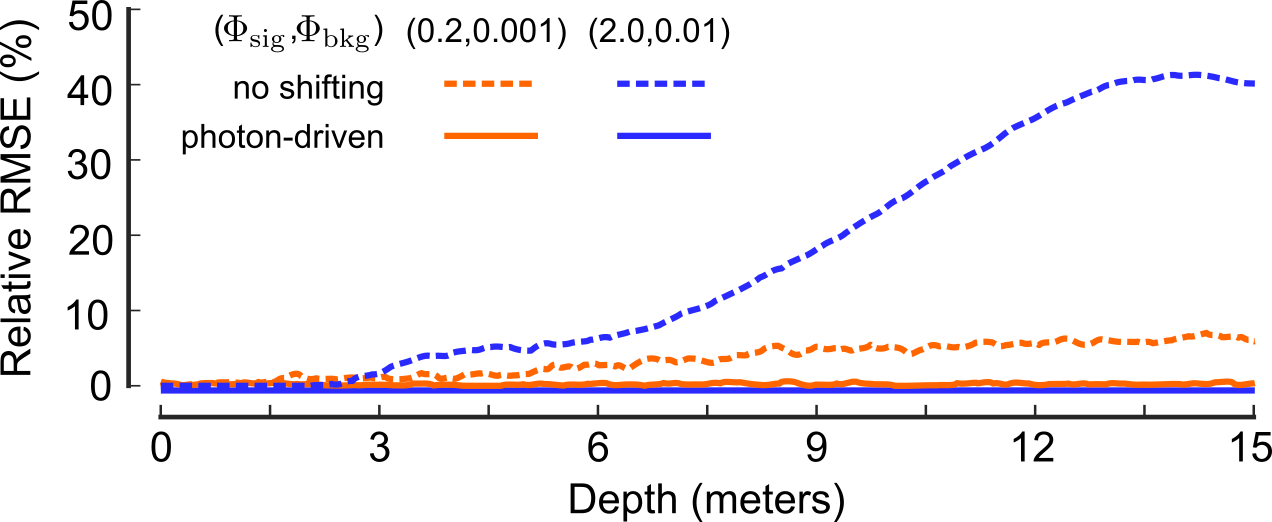}
  \caption{Asynchronous acquisition schemes minimize overall depth error, and
  as a byproduct also provide uniform error with depth. The performance gain is
  larger at longer distances.}
  \label{fig:suppl_rmse_vs_depth}
\end{figure}

Suppl. Fig.~\ref{fig:suppl_rmse_vs_depth} shows error vs. depth for photon-driven and
conventional synchronous acquisition for different source and background flux
levels. There are two key observations:
\begin{enumerate}

\item The performance gain from photon-driven acquisition is greater at longer
  distances. (Although not shown here, similar conclusion can be drawn for
  deterministic shifting as well.)

\item RMSE is approximately depth-invariant for photon-driven shifting.
\end{enumerate}

Achieving uniform error over all depths is not our primary goal, but a
byproduct of our optimal acquisition schemes. We define optimality in terms of
a surrogate of $\ell_2$ depth error metric which measures overall depth error.
Result~\ref{thm1} shows that the optimal acquisition scheme that minimizes our error
metric also has a constant denominator sequence for all depths. 

\clearpage\newpage
\section{Asynchronous Shifting: Practitioner's View\label{sn:free_vs_det}}
In this supplementary note we provide practical guidelines for design of
asynchronous acquisition strategies. We provide a comparison between
photon-driven shifting and uniform shifting with optimal active time
$m^\text{opt}$ and show that there are certain regimes where photon-driven
shifting has slightly worse error performance than uniform shifting.  We also
analyze the effect of number of SPAD cycles on depth error performance of
photon-driven acquistion.

\subsection*{Derivation of Expected Denominator Sequence for Photon-Driven Acquisition}
We consider a low SBR scenario with
$r_i \approx \lam$ for all $i$. Let $(D_i)_{i=1}^B$ denote the
denominator sequence obtained over $L$ SPAD cycles and a
fixed acquisition time $T$. Since the length of each SPAD cycle is random, the
number of cycles in time $T$ is also random. By the equivalence of the length
of the active period of a cycle and its contribution to the total denominator,
we have:
\[
  D_1 + D_2 + ... + D_B = M_1 + M_2 + ... + M_L
\]
where $M_i$ is a random variable denoting the length of the $i^\text{th}$
measurement window. Note that the $M_i$'s are i.i.d. with a discrete
exponential distribution:
\[
  P(M_i = k) = e^{-(k-1)\lam} \left(1 - e^{-\lam}\right)
\]
for $k = 1,2,3,\ldots.$ Therefore, $\E[M_i] = \frac{1}{1 -
e^{-\lam}}$. Taking expectation on both sides of the above equation, we
get:

\begin{align}
\label{sum_identity_1} \mathbb{E}[D_1 + D_2 + ... + D_B] &= \mathbb{E}[M_1 + M_2 + ... + M_L] \\
\label{sum_identity_2} &\approx \mathbb{E}[L] \mathbb{E}[M_i] = \frac{\mathbb{E}[L]}{1 - e^{-\lam}} \,,
\end{align}
where, we have assumed that $M_i$ and $L$ are approximately independent.  We
also have:
\[
  (M_1 + \td) + (M_2 + \td) + ... + (M_L + \td) = T \,.
\]
Taking the expectation on both sides and using Eq.~(\ref{sum_identity_2}), we get:
\[
  \frac{\mathbb{E}[L]}{1 - e^{-\lam}} + \mathbb{E}[L] \td = T
\]
which implies that
\[
  \mathbb{E}[L] = \frac{T}{\frac{1}{1 - e^{-\lam}} + \td} \,.
\]
Combining with Eq.~(\ref{sum_identity_1}), we get:
\[ 
  \mathbb{E}[D_1+D_2 + ... + D_B] = \frac{T}{1 + (1 - e^{-\lam}) \td} \,.
\]
From Result~\ref{thm3}, we know that $\E[D_i]$ is constant $\forall\,i$ as $L
\to \infty$.  Under this assumption, we have for $1 \leq i \leq B$:
\begin{equation}
  \E[D_i]|_\text{photon-driven} = \frac{T}{B(1 + (1 - e^{-\lam}) \td)}. \label{eq:expected_den_photon_driven}
\end{equation}

\noindent{\bf Comparing expected denominator sequences for uniform shifting and photon-driven shifting:}
Recall from \ref{sn:mopt_derivation} the expected denominator sequence for uniform shifting is given by: 
\begin{equation}
  \E[D_i]|_\text{uniform} = \frac{T(1 - e^{-m \lam})}{B(m\Delta + \td)(1 - e^{-\lam})}. \label{eq:expected_den_deterministic}
\end{equation}

% Figure shows the
% optimal $D_i^m$ and $D_I^{\text{free}}$ as a function of $\lambda$, for
% different values of $\tau$. For large $\tau \lambda$, they are similar, which
% is expected since we saw earlier that deterministic shifting is able to achieve
% both sides of the trade-off for such conditions. For low $\tau \lambda$,
% however, free running is much higher, highlighting the limited capability of
% deterministic shifting in straddling the trade-off. The right sub-figure shows
% error curves for the same settings as the denominator curves on the left.
% Clearly, the regions where free running outperforms deterministic shifting
% match the predictions from the denominator curves. Note that the error curves
% are not monotonic with flux. This effect is discussed in the next section.

\begin{figure}
  \centering\includegraphics[width=0.9\columnwidth]{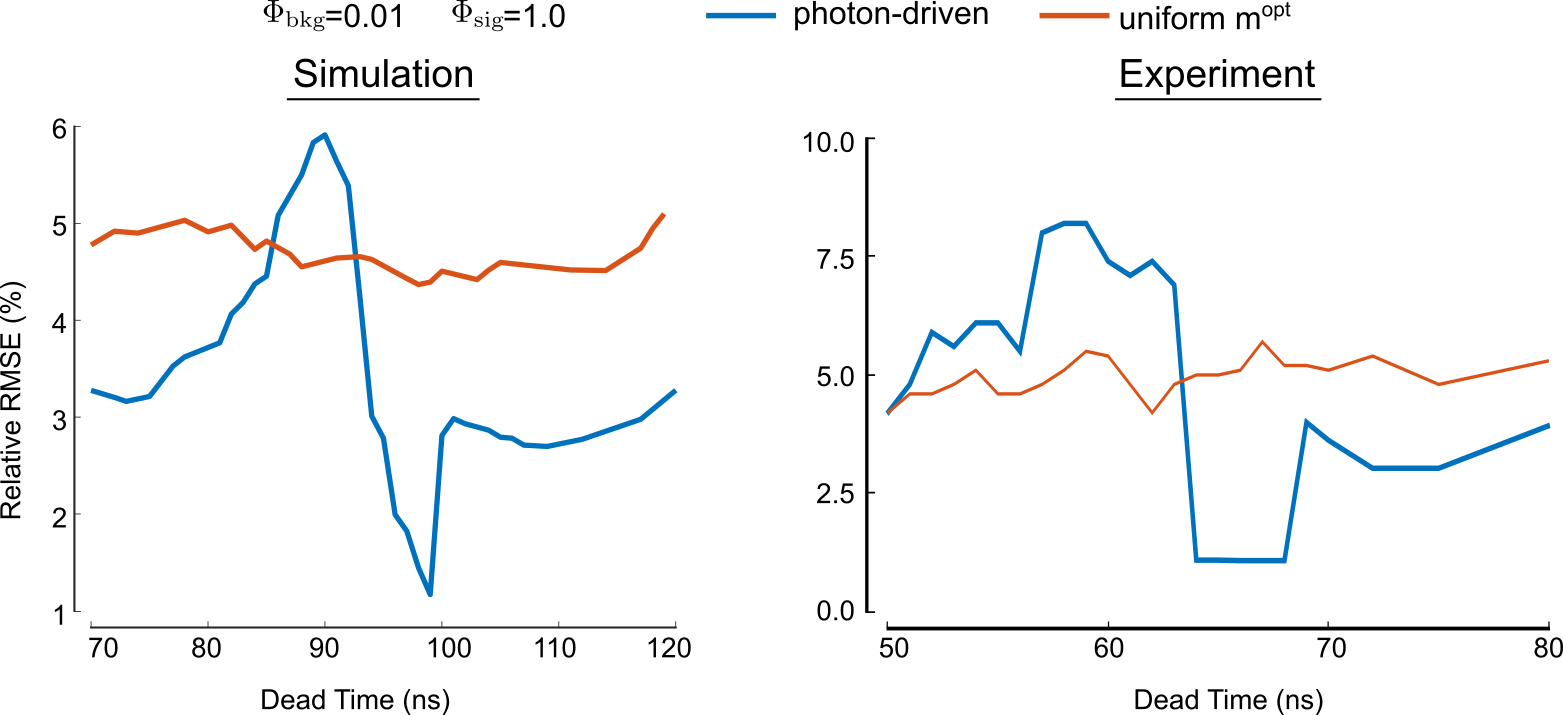}
\caption{In some special scenarios, depending on dead time and incident flux, uniform shifting may outperform photon-driven shifting.
\label{fig:det_better_opt}}
\end{figure}

Observe that $\E[D_i]|_\text{photon-driven} > \E[D_i]|_\text{uniform}$ for all
$m \geq 1$, $\lam$ and $\td$.  To a first order approximation, the expected
denominator sequence alone determines the depth estimation performance of the
various techniques that we discussed. This suggests that photon-driven shifting
should be always better than uniform shifting. However, in some cases,
depending on the dead time and incident flux levels, deterministic shifting can
outperform photon-driven acquisition. One such scenario is demonstrated in
Suppl. Fig.~\ref{fig:det_better_opt} using both simulation and experimental
results.  At certain values of dead time, photon-driven shifting fails to
achieve uniform shifts in the fixed acquistion time.

%However, in certain scenarios,
%we also need to consider the variance of the denominators. This is especially
%important for free running, where the variance can be very high for certain
%combinations of $\lambda$ and $\tau$, leading to higher errors than expected.
%See the appendix for a more detailed discussion.
%Fig.~\cite{fig:timing_diagram} shows the simulated depth errors for uniform
%shifting and photon-driven shifting, as a function of the dead time $\td$.
%Clearly the free running surface lies below the deterministic surface. At
%$\tau$ around $900$, which is just less than $n = 1000$, the free running error
%creeps above the deterministic error slightly. This is again because the
%denominator variance is large in this

\paragraph{Effect of reducing laser cycles on photon-driven shifting}
\begin{figure}[!ht]
  \centering\includegraphics[width=0.5\columnwidth]{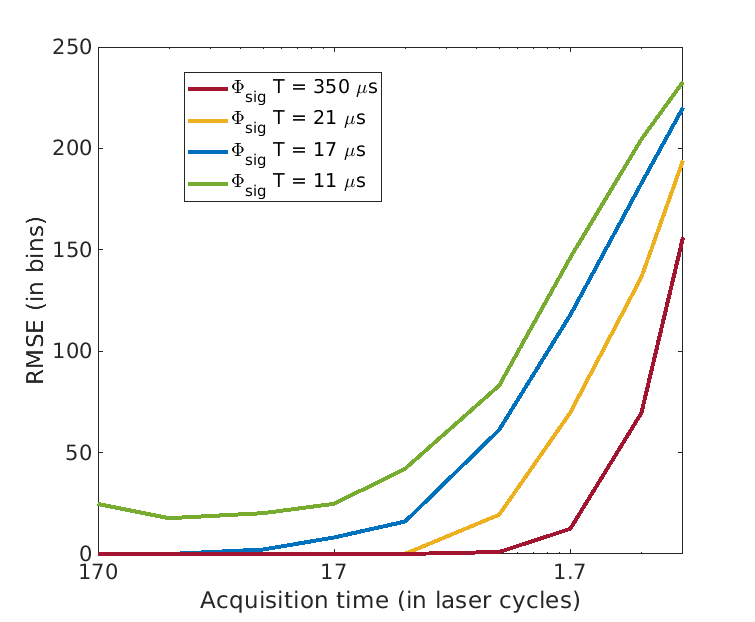}
  \caption{ {\bf Effect of reducing laser cycles on photon-driven shifting with optimal attenuation.}
  Each curve has a fixed $\muu T$ product. As $T$ decreases and $\muu$ increases, the RMSE rises. The effect of decreasing $T$ can be overcome to some extent by increasing $\muu$. 
  \label{fig:effect_of_T}}
\end{figure}

% Given various shifting and filtering
% pile up mitigation analyzed in this paper and computational strategies described
% elsewhere in literature, the practitioner may find it overwhelming to choose
% one method over the other. What is the bottom line? Which method is the best?
% We have already established that photon-driven shifting with optimal attenuation has the best performance in the space of asynchronous acquisition schemes.
Intuitively, one would expect asynchronous acquisition techniques to require a
large number of cycles, in order to achieve uniform shifting and remove pileup
as motivated in the main text. However, it turns out that our combined method
of photon-driven shifting and optimal attenuation can work with very few laser
cycles, provided that the signal flux is high. This is true even for relatively
high ambient flux levels (we consider $\lam = 0.05$). Intuitively, even though
uniform shifting cannot occur with laser cycles much less than the number of
bins $B$, the combination of high effective $\muu$ and low effective $\lam$
obviates the need for shifting.

In Suppl. Fig.~\ref{fig:effect_of_T}, we evaluate depth error performance as number of
laser cycles is decreased, while keeping the $\muu T$ product constant. For all
$\muu$ levels, RMSE curves have flat valleys, with errors starting to rise
beyond a certain $T$ and reaching the maximum possible error eventually. It is
interesting to see the transition from the valley to max error for different
$\muu$ levels. As the $\muu$ level is increased, the transition happens at a
lower acquisition time, until $T$ becomes less than about $2$ laser cycles.
Beyond this, the error cannot be reduced no matter how high $\muu$ is made. In
some sense, this is the limiting point of our method.

\clearpage\newpage
\section{Combining Shifting Strategies with Optimal Flux Attenuation\label{sn:filtering_with_shifting}}
In this section we derive an expression for optimal flux attenuation for
photon-driven acquisition. We also provide some information theoretic arguments
for why it is important to combine asynchronous acquisition schemes with flux
attenuation to achieve optimal depth error performance.

\subsection*{Optimal Attenuation Fraction for Photon-Driven Shifting}
We will use similar techniques as the derivation of Result~\ref{thm1} from \ref{sn:res1_res2_proofs}
and introduce an additional variable $\Upsilon$ for flux attenuation fraction. In particular
the expected denominator sequence in Eq.~(\ref{eq:expected_den_photon_driven}) becomes:
\begin{equation}
  \E[D_i] = \frac{T}{B(1 + (1 - e^{-\Upsilon\lam}) \td)}. \label{eq:expected_den_photon_driven_upsilon}
\end{equation}
From Eq.~(\ref{eq:upperbound}), we have:
\begin{align*}
 \frac{1}{B} \sum_{\tau=1}^B \P(\widehat \tau \neq \tau)  &\leq \frac{1}{B}\sum_{i,\tau=1}^B \frac{1}{2}\exp{\left( - \frac{1}{2}\frac{(q_\tau - q_i)^2}{ \frac{q_i (1 - q_i)}{\E[D_i]} + \frac{q_{\tau} (1 - q_{\tau})}{\E[D_{\tau}]}} \right)} \\
 &\approx \frac{1}{B}\sum_{i,\tau=1}^B \frac{1}{2}\exp{\left( - \frac{1}{2}{{\E[D_{i}]} (q_\tau - q_i)} \right)} \\
 &= \frac{B}{2}\exp{\left( - \frac{1}{2}{{\E[D_{i}]} (q_\tau - q_i)} \right)}
\end{align*}
where in the second step, we have assumed that the denominator sequence is uniform, $q_i, q_\tau \ll 1$ and $q_{\tau} - q_i \approx q_\tau + q_i$.
Substituting Eq.~(\ref{eq:expected_den_photon_driven_upsilon}), the optimal attenuation fraction is given by:
\begin{align*}
  \Upsilon^{\text{opt}} &= \argmax_{\Upsilon} \left[{\E[D_{i}]} (q_\tau - q_i)\right] \\
&= \argmax_{\Upsilon} \frac{T e^{-\Upsilon \lam}(1 - e^{-\Upsilon \muu})}{B(1 + (1 - e^{-\Upsilon \lam}) t_d)}.
\end{align*}

The optimal attenuation fraction $\Upsilon^{\text{opt}}$ depends on various system parameters (number of bins, acquisition time and the dead time), which are known. $\Upsilon^{\text{opt}}$ also depends on the signal and background flux levels, which can be estimated a priori using a small number of laser cycles. Using the estimated values, the optimal flux attenuation can be determined, which can be used to further reduce the pile-up, and thus improve the depth estimation performance. 

Fig. 6 in the main paper shows simulation results of depth RMSE for the conventional synchronous mode and photon-driven shifting over a range of attenuation factors and two different dead times. The locations of the minima agree with the theoretical expression derived above. There are two key observations that can be made from the figure. First, the optimal attenuation fraction with shifting is much higher than that for conventional synchronous acquisition. Second, combining
attenuation with photon-driven shifting can provide a large gain in depth error performance, reducing the RMSE to almost zero for some combinations of SBR and dead time.

\subsection*{Information Theoretic Argument for Combining Shifting with Flux Attenuation} 

%The idea of shifting helped reduce depth error by increasing the minimum
%expected denominator for a depth bin. Optimal shifting (both deterministic and
%photon driven) further reduced error by maximizing the total denominator. Now
%we raise the question: is there a way to improve depth estimation other than by
%increasing the denominator sequence?
%
%We can think about increasing $D_i$ as increasing the number of samples that we
%get to estimate the unknown flux $r_i$. But along with the \emph{quantity} of
%data, we also need to consider the \emph{quality} of data, i.e., the value that
%we get per sample. It turns out that there is a counterintuitive way to
%increase that value itself: attenuating the flux. In other words, scaling the
%unknown flux by some known factor $< 1$ can help us estimate the flux better.
%
%It is not hard to see why this can be true. Consider a situation with a high
%flux of $r_i = 10$ photons per bin. In this case, the probability of photon
%detection is $1 - e^{-10} \approx 1$. Therefore, with high probability, each
%sample of the bin will yield a photon detection i.e. $N_i = D_i$. Now consider
%another situation with $r_i = 20$. In this case too, $N_i = D_i$ with near 1
%probability. Clearly it is impossible to tell apart a flux of $10$ or $20$ from
%the data. Now if we scale the flux down to $0.5$ and $1$, the problem becomes
%easier.

We formalize the notion of discriminative power of the histogram
data for distinguishing between incident flux waveform levels using a concept
from statistics called \emph{Fisher information} [Suppl.~Ref.~\ref{ref_kay},
pp.~35].  This concept was also used in [Suppl.~Ref.~\ref{ref_rapp}]. Fisher
information measures of the rate of change of the likelihood function with
respect to the unknown parameters $r_i$. For any attenuation coefficient
$\Upsilon$, we can compute:
\begin{align*}
    \mathcal{I}(r_i; \Upsilon) = \E{\left[\left( \frac{\partial}{\partial r_i} \log \mathcal{L}(N_i; D_i, r_i, \Upsilon)\right)^2\right]}
\end{align*}
Substituting the expression for likelihood from Eq.~(\ref{eq:likelihood}) and simplifying yields: 
\begin{align}
\label{fisher}
    \mathcal{I}(r_i; \Upsilon) = \frac{D_i \Upsilon^2}{e^{\Upsilon r_i} - 1}
\end{align}
Asynchronous acquisition methods can only change $D_i$ which appears in the
numerator of the Fisher information. Attenuation can further increase the
Fisher information and lead to better flux utilization than would be possible
with shifting alone.

\clearpage\newpage
\section{Details of Simulations and Experiments \label{sn:sim_expt_details}}

\subsection*{Calculating RMSE}
We calculate RMSE in a modulo-B sense in our simulations and experiments. For
example, this means that the first and the last histogram bins are considered
to be only $1$ bin apart. Let $M$ be the number of Monte Carlo runs, $B$ be the
number of histogram bins, $\tau_i$ be the true depth bin index and $\widehat\tau_i$
be the estimated depth bin index for the $i^\text{th}$ Monte Carlo run. We
calculate RMSE using the following formula:

\[
  \mathsf{RMSE} = \left [ \frac{1}{M}\sum_{i=1}^{M}
  \left(\frac{B}{2} - \left( \widehat{\tau}_i - \tau_i + \frac{B}{2}\right)\, \mathsf{mod}\, B  \right)^2 \right]^{\frac{1}{2}}.
\]

\subsection*{Monte Carlo Simulations}
In Figs.~\ref{fig:tcspc_vs_determinsitic} and \ref{fig:photon_driven_surfaceplot}, the dead time was $\SI{10}{\nano\second}$, and the exposure time was $\SI{2.5}{\micro\second}$
In Fig.~\ref{fig:error_vs_m}, the exposure times were \SI{2.5}{\micro\second}
and \SI{2.9}{\micro\second} respectively for the two dead times, to acquire an
equal number of SPAD cycles in both cases.
At low incident flux and longer dead times, an active time duration longer than
$B$ may be needed to increase the probability that the SPAD captures a photon.

\subsection*{Experimental Setup}
\begin{figure}[!ht]
  \centering \includegraphics[width=0.9\textwidth]{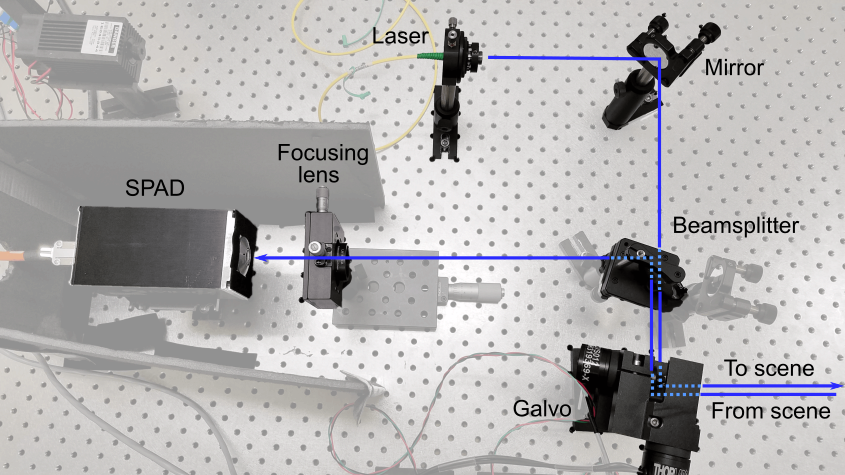}
  \caption{{\bf Experimental setup} The setup consists of a collimated pulsed
  laser and a single-pixel SPAD detector optically collocated using a
  beamsplitter. A pair of galvo mirrors scans the image plane and the light
  returning from the scene is focused on the detector using a focusing lens.
  The detector and focusing lens are enclosed in a light-tight box (not shown
  here) so that the all the light arriving at the detector pixel must pass
  through the lens. \label{fig:setup}}
\end{figure}

Suppl.~Fig.~\ref{fig:setup} shows various components of our experimental setup.
The SPAD is operated by a programmable control unit and photon timestamps are
captured by a TCSPC module (not shown).

% \clearpage\newpage
\subsection*{Single-Pixel Experiment Results}
\begin{figure}[!t]
  \centering\includegraphics[width=0.9\columnwidth]{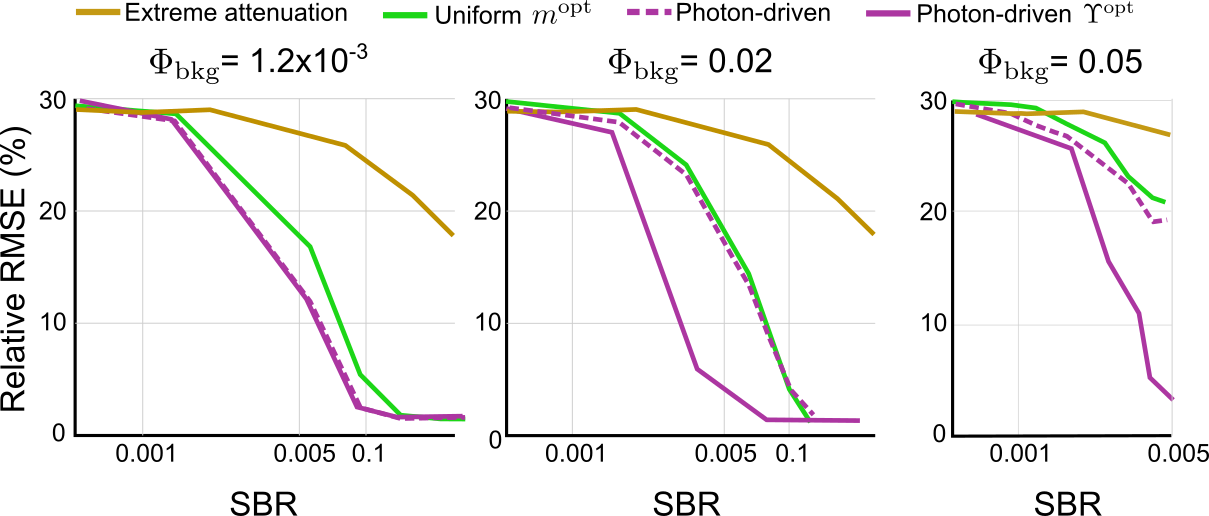}
  \caption{{\bf RMSE in single-pixel experiments.} 
  Asynchronous schemes outperform the state-of-the-art method of synchronous
  acquisition with extreme attenuation at all flux levels. A combination of
  photon-driven shifting with optimal attenuation provides an order of
  magnitude lower RMSE at high ambient flux levels.\label{fig:single_pixel_expt}}
  \vspace{-0.15in}
\end{figure}
Suppl. Fig.~\ref{fig:single_pixel_expt} shows
depth RMSE at four different ambient flux levels for a range of SBR values. A
combination of photon-driven shifting with optimal attenuation provides the
best performance of all techniques. For high ambient flux levels, even at high
SBR, the conventional 5\% rule-of-thumb and shifting alone fail to provide
acceptable depth reconstruction performance. This suggests that the optimal
acquisition strategy for SPAD-based LiDAR must use a combination of both
shifting and attenuation.

\clearpage\newpage
\subsection*{Additional Result Showing Effect of Number of SPAD Cycles}
\begin{figure}[!ht]
  \centering \includegraphics[width=0.9\textwidth]{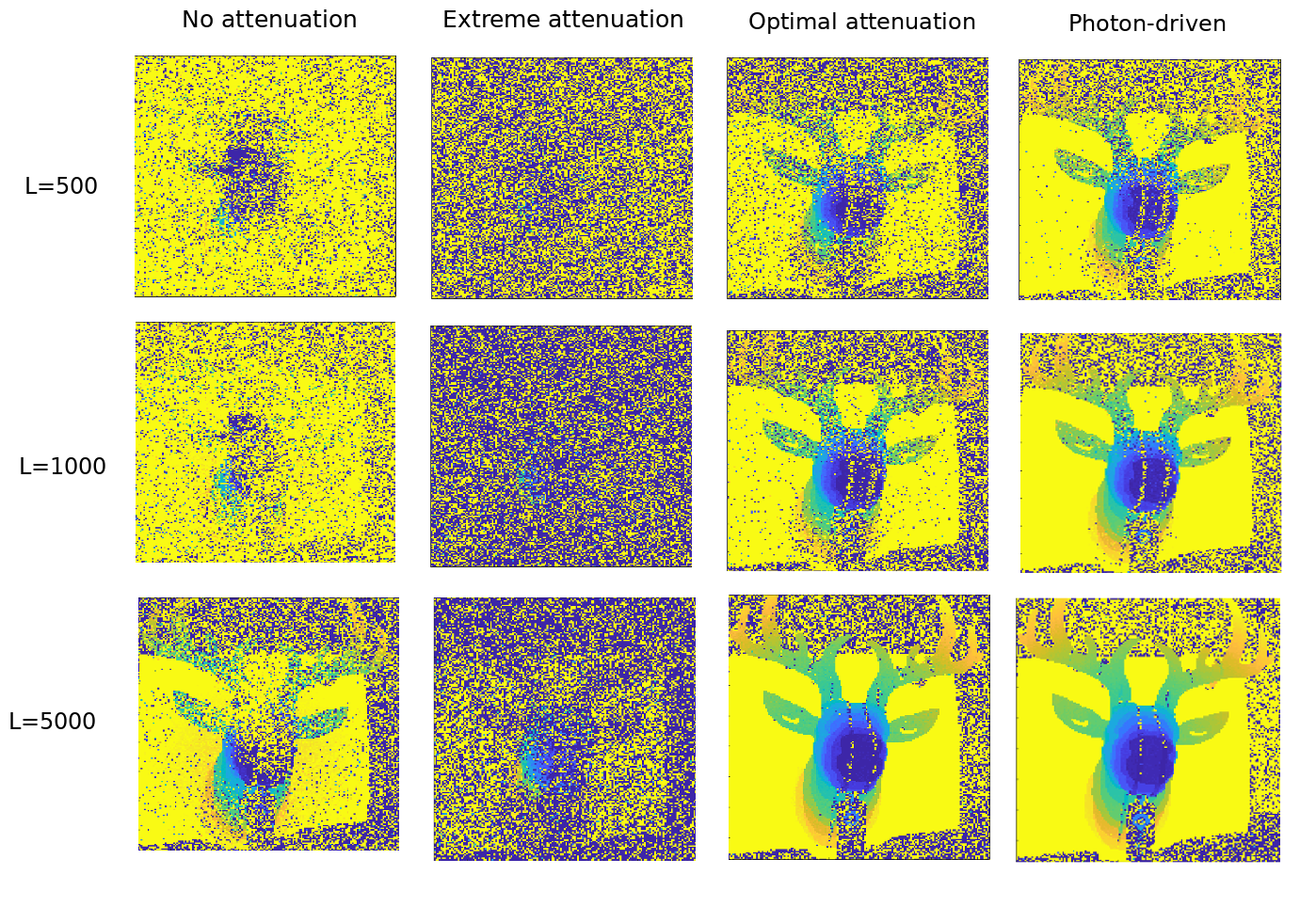}
  \caption{{\bf Depth reconstructions for varying number of SPAD cycles for
  ``Reindeer'' scene.} Observe that the reconstruction accuracy improves as the
    number of cycles increases. Photon-driven shifting provides better
    reconstruction performance than all synchronous acquisition schemes. The
    source flux $\muu=0.5$ and ambient flux $\lam=0.01$ for this experiment which
    corresponds to an SBR of 50.}
\end{figure}

\clearpage\newpage

\section*{Supplementary References}
\renewcommand*\labelenumi{[\theenumi]}
\begin{enumerate}
    \item K. Murphy, Machine Learning: A Probabilistic Perspective.
      Cambridge, MA: MIT Press, 2012, pp. 35. \label{ref_murphy}
    \item C. Daskalakis, G. Kamath, and C. Tzamos. On the structure,
      covering, and learning of poisson multinomial distributions.
      arXiv preprint, 2015. \label{ref_pmd}
    \item A. Gupta, A. Ingle, A. Velten, and M. Gupta. Photon flooded
      single-photon 3d cameras. arXiv preprint arXiv:1903.08347,
      2019. \label{ref_gupta}
    \item S. Kay, Fundamentals of Statistical Signal Processing: Estimation
      Theory. Upper Saddle River, NJ: Prentice Hall, 1993, pp.~173--174.
      \label{ref_kay}
    \item R. M. Corless, G. H. Gonnet, D. E. G. Hare, D. J. Jeffrey,
      and D. E. Knuth. On the LambertW function. Advances in
      Computational Mathematics, 5(1):329–359, Dec 1996. \label{ref_lambertw}
    \item J. Rapp, Y. Ma, R. Dawson, and V. K. Goyal. Dead
      time compensation for high-flux ranging. arXiv preprint
      arXiv:1810.11145, 2018. \label{ref_rapp}
    \item G. Grimmett, D. Stirzaker, Probability and Random Processes.
      Oxford, UK: Oxford University Press, 2001, pp.~227. \label{ref_grimmett}
\end{enumerate}

\clearpage

\end{document}